\documentclass[10pt,journal]{IEEEtran}
\newtheorem{theorem}{Theorem}

\newtheorem{lemma}{Lemma}
\newtheorem{definition}{Definition}
\newtheorem{remark}{Remark}

\newtheorem{assumption}{Assumption}

\newtheorem{proposition}{Proposition}
\usepackage{graphicx}
\usepackage{subfigure}
\usepackage{array} 
\usepackage{amsmath}
\usepackage{amssymb}
\usepackage{mathtools} 
\usepackage{latexsym}
\usepackage{mathrsfs}
\usepackage{cite}
\usepackage{enumerate} 
\usepackage{float}
\usepackage{amsfonts} 
\usepackage{epsfig} 
\usepackage{epstopdf}
\usepackage{balance} 
\usepackage[dvipsnames]{xcolor}
\usepackage{tikz}\newcommand*{\circled}[1]{\lower.7ex\hbox{\tikz\draw (0pt, 0pt)%
		circle (.4em) node {\makebox[1em][c]{\small #1}};}}
\usepackage{color}
\usepackage[scaled]{helvet}
\usepackage{tikz,xcolor,hyperref} 
\definecolor{lime}{HTML}{A6CE39}
\DeclareRobustCommand{\orcidicon}{\begin{tikzpicture}\draw[lime, fill=lime] (0,0) circle [radius=0.16] node[white] {{\fontfamily{qag}\selectfont \tiny ID}}; \draw[white, fill=white] (-0.0625,0.095) circle [radius=0.007];  \end{tikzpicture}\hspace{-2mm}}
\foreach \x in {A, ..., Z}{\expandafter\xdef\csname orcid\x\endcsname{\noexpand\href{https://orcid.org/\csname orcidauthor\x\endcsname}{\noexpand\orcidicon}}}

\begin{document}
\title{\textcolor{black}{Distributed Matrix Pencil Formulations for Prescribed-Time Leader-Following Consensus} of MASs with Unknown Sensor Sensitivity}
\author{Hefu Ye, Changyun Wen, \IEEEmembership{Fellow,~IEEE}, and Yongduan Song$^{*}$, \IEEEmembership{Fellow,~IEEE} 
%\thanks{This work was supported in part by the National Key Research and Development Program of China under Grant 2022YFB4701400/4701401, and in part by the National Natural Science Foundation of China under Grant 61933012, Grant 61991400, Grant 61991403, Grant 62250710167, Grant 61860206008,  Grant 62273064. {\it (Corresponding author: Yongduan Song.)}}
\thanks{Hefu  Ye  and Yongduan Song are   with  the School of Automation, Chongqing University, 400044, China. Engineering, Nanyang Technological University,  639798, Singapore. (e-mail: yehefu@cqu.edu.cn; ydsong@cqu.edu.cn)}
\thanks{Changyun Wen is with the School of Electrical and Electronic Engineering, Nanyang Technological University,  639798, Singapore. (e-mail:  ecywen@ntu.edu.sg).} 
}

\maketitle

\begin{abstract}  
In this paper, we address the problem of prescribed-time leader-following consensus of heterogeneous multi-agent systems (MASs) in the presence of unknown sensor sensitivity. Under a connected undirected topology, we propose a time-varying dual observer/controller design framework that makes use of regular local and inaccurate feedback to achieve consensus tracking within a prescribed time. In particular, the developed analysis framework is applicable to MASs equipped with sensors of different sensitivities. 
\textcolor{black}{One of the design innovations involves constructing a distributed matrix pencil formulation based on worst-case  sensors, yielding control parameters with sufficient robustness yet relatively low conservatism.  Another novelty is the construction of the control gains, which  consists of the product of a proportional coefficient obtained from the matrix pencil formulation and a classic time-varying function that grows to infinity or a novel bounded time-varying function.}
Furthermore, it is possible to extend the prescribed-time distributed protocol to infinite time domain by introducing the bounded time-varying gain technique without sacrificing the ultimate control accuracy, and the corresponding technical proof is comprehensive.  The effectiveness of the method is demonstrated through a group of 5 single-link robot manipulators.
\end{abstract}

\section{Introduction}  
%\textcolor{black}{Different from autonomous finite-time and fixed-time controllers, the prescribed-time controller is an non-autonomous  (time-varying)  controller that ensures that the system state will be regulated to the origin within a pre-specified time interval, regardless of the initial state and any other design parameter.} 

\IEEEPARstart{I}N recent years, \textcolor{black}{there has been a notable surge of interest in the field of prescribed-time control, which allows arbitrarily prescribe the finite convergence time of a closed-loop system regardless of initial conditions and any other design parameter.} It is wroth noting that the idea of prescribed-time control can be traced back to the proportional navigation in tactical missile guidance \cite{Zarchan07}, and early explorations of prescribed-time stability primarily revolved around the optimal control of linear systems \cite{1964-Sidar-ptc}. %One of the distinguishing features of these works is that the controllers developed therein are non-autonomous, which sets them apart from the so-called finite-/fixed- and predefined-time controllers, since the latter are all autonomous, see e.g., \cite{2018-Sanchez-predefined-time} and surveys on prescribed-time control\cite{2023-Song-survey}. 
Prescribed-time control for nonlinear systems was thoroughly developed in \cite{2017-Auto-Song}, and since then, several appealing techniques have been introduced for achieving prescribed-time stabilization for complex single-plant nonlinear systems. These methods include dynamic high-gain feedback\cite{2020-NYU-dynamic-high-gain,2021-NYU-ACSP,2021-RNC-NYU}, bounded linear time-varying feedback\cite{2021-zhoubin-tac-prescribed}, time-varying and non-Lipschitz synergistic feedback\cite{2022-orlov-auto}, among others. An objective fact is that the current research on prescribed-time control for MASs  is primarily focused on linear systems. For instance, \textit{Wang et al.} in \cite{2018-Auto-Wang} and \cite{2018-Cyber-Wang} developed distributed prescribed-time containment and consensus protocols for single and $n$-th order integrators, respectively. \textit{Kan et al.} in \cite{2017-DSMC-Kan} and \textit{Yucelen et al.} \cite{2019-TAC-Yucelen} established a prescribed-time consensus framework for linear MASs based upon a time-axis transformation. %With the help of this framework, some typical robust approaches were studied for MASs with uncertainties in \cite{2021-TCNS-Arabi,2020-SCL-Yucelen,2021-RNC-Arabi}. 
\textcolor{black}{In addition, distributed finite-time and predefined-time protocols were constructed for MASs with directed communication in  \cite{2021-tnnls-wangqing} and \cite{2021-wangqing}, respectively. 
Recently, based upon the solution of a parametric
Lyapunov equation, fully distributed state-feedback protocols were developed to solve the problems of asymptotic and prescribed-time  consensus of linear MASs in \cite{2023-scl-zhangkai} and \cite{2023-scis-zhangkai}, respectively.}
In contrast to the aforementioned consensus results where only cooperative communication exist among agents, \textit{Ning et al.} in \cite{2020-TAC-Hanqinglong} introduced a prescribed-time bipartite consensus protocol for double integrators where antagonism exists in the communication. Later on, \textit{Chen et al.} extended the prescribed-time bipartite consensus result to single integrators with event-triggered communications in \cite{2020-Chen-Prescribed-event}.

A significant challenge occurs in distributed prescribed-time consensus tracking when the considered system model becomes nonlinear and contains uncertainties. This is particularly true in cases where only inaccurate output measurements of each agent are available. Based upon accurate measurement signals, some preliminary efforts on prescribed-time output-feedback control for nonlinear MASs have been made in recent studies. In \cite{2023-Cyber-Zou}, a practical prescribed-time control approach was proposed for second-order leader-following systems. In \cite{2022-RNC-Ke}, prescribed-time precise consensus was achieved for second-order disturbed MASs based upon a mismatched disturbance observer. In \cite{2020-SCL-Chenxiandong}, \textit{Chen et al.} made remarkable progress in the prescribed-time control of high-order MASs in strict-feedback form by modifying feedback domination design approach. To the best of the authors' knowledge, research on distributed prescribed-time consensus of heterogeneous MASs in the presence of unknown measurement sensitivity remains limited so far. On the one hand,  it is not straightforward to develop a stability analysis framework suitable for inaccurate measurements, especially when each agent is equipped with sensors of different sensitivities. On the other hand, the aforementioned distributed methods are invalid beyond the prescribed time interval since the control gain escapes to infinity as $t$ approaches the terminal time. It is therefore of theoretical significance and practical importance to develop a modified scheme involving bounded gains that drives the closed-loop system not to the destination, but to its neighborhood within a prescribed time, and ultimately to the origin.

In the present work, the high-gain domination approach is revisited for prescribed-time distributed control of heterogeneous leader-following MASs. Unlike classical literature that relies on high-gain feedback and can only accommodate uncertainties with known growth rates, the system  under consideration allows for uncertainties with unknown growth rates and coupled unmeasurable states. In addition, inspired by the recent works of \textit{Krishnamurthy and  Khorrami}\cite{2020-NYU-IFAC,2020-CDC-matrix-pencil,2022-scl-matrix-pencil,2023-tac-matrix-pencil}, \textcolor{black}{we robustify the standard matrix pencil-based design procedure in the presence of multiplicative measurement noise and propose a class of novel distributed matrix pencil formulations for the prescribed-time control design.} The basic design idea is to convert the problem of prescribed-time leader-following consensus into the asymptotic stabilization of related errors with the help of state scaling and time-axis transformation, and then cast the Lyapunov inequality into an equivalent matrix pencil form to solve for suitable design parameters. %%%Besides, the observer and controller are designed with a dual architecture, which makes the resulting controller simple in structure, does not involve recursive calculations, and does not have the peaking phenomena. 
Particularly, the most conservative sensitivity parameters of sensors are used as a benchmark for designing controller parameters to ensure that the resulting algorithm is effective for all agents equipped with different sensors. The stability analysis is completed in an elegant way that once the boundedness of the Lyapunov function is obtained, the boundedness of all internal signals including inputs and the prescribed-time convergence of  tracking errors can be obtained.  \textcolor{black}{The contributions of this article are summarized as follows:}
\begin{itemize}
	\item[$1)$] \textcolor{black}{Prescribed-time leader-following consensus is achieved for heterogeneous  MASs by employing the time-varying feedback and high-gain design, and the corresponding practical prescribed-time solution is then formulated via a modified bounded time-varying feedback}; %新方法实现PT，且提出了修正策略
	\item[$2)$] \textcolor{black}{The mathematical expression of agent-based sensor uncertainty is constructed, and the philosophy of using high-gain design to solve the uncertainty of different sensors is given based on the worst sensor performance}; %建立了处理不同传感器故障的哲学 
	\item[$3)$] \textcolor{black}{A distributed matrix pencil formulation that is robust to multiplicative noise is proposed to reduce the conservatism of control parameters.}  %提出了分布式矩阵束公式
\end{itemize}

\section{Problem formulation and preliminaries} \label{sec-problem-statement}
We first introduce some notations and definitions. Given a vector $q=[q_1,\dots,q_n]^{\top}$, $q^{\top}$ and $|q|$ denote  the transpose  and the Euclidean norm of the vector $q$, respectively,  the notation $|q|_e$ is used to denote the vector of element-wise magnitudes $[|q_1|,\dots,|q_n|]^{\top}$, and the relation $|q|_e \leq_e |x|_e$ indicates the set of element-wise inequalities $|q_i|_e\leq |x_i|_e$, where $x$ is a vector (or a matrix) of the same dimension as $q$.   Given a matrix $\mathbf{Q}$ {(where bold notations denote matrices)}, 
$\det(\mathbf{Q})$ denotes the determinant of matrix $\mathbf{Q}$, $|\mathbf{Q}|$ denotes the  Frobenius norm of matrix $\mathbf{Q}$,  $|\mathbf{Q}|_e$ denotes the matrix of the same dimensions with each element being the magnitude of the corresponding element of $\mathbf{Q}$,  $\lambda_{\max}(\mathbf{Q})$ and $\lambda_{\min}(\mathbf{Q})$ denote its maximum and minimum eigenvalues, respectively, and   $\operatorname{diag}\{\mathbf{Q}\}$ denotes the diagonal matrix whose diagonal elements are equal to the diagonal elements of $\mathbf{Q}$. The diagonal matrix with diagonal elements $q_1,\cdots,q_n$ is denoted by $\operatorname{diag}\{q_1,\cdots,q_n\}$. If $q_1,\cdots,q_n$ are matrices, $\operatorname{diag}\{q_1,\cdots,q_n\}$ denotes the block diagonal matrix with blocks on the principal diagonal being $q_1,\cdots,q_n$.

Consider a multi-agent system comprising a leader and $N$ agents. The dynamics of each agent are described by:
\begin{equation}\label{model}
\begin{aligned}
\dot{x}_k&=\mathbf{A}x_k+Bu_k+F_k(t,x_k), \\
y_k&=\textcolor{black}{\theta_k(t)}Cx_{k},~k\in \{0,1,\cdots,N\},
\end{aligned}
\end{equation} 
\begin{equation*}  
\mathbf{A} =\left[\begin{array}{cc}
\hspace{-0.05cm}	0 &  \mathbf{I}_{n-1} \hspace{-0.06cm}  \\
\hspace{-0.05cm}	0 & 0   \hspace{-0.06cm}
\end{array}\right] ,~B=\left[\begin{array}{c}
\hspace{-0.05cm}	0_{(n-1)\times 1} \hspace{-0.06cm}       \\
\hspace{-0.05cm}	1    \hspace{-0.06cm}
\end{array}\right], ~C=\left[\begin{array}{c}
\hspace{-0.05cm}	1   \hspace{-0.06cm}  \\
\hspace{-0.05cm}	0_{(n-1)\times 1 \hspace{-0.06cm} } 
\end{array}\right]^{\top}\hspace{-0.1cm},
\end{equation*}
where $x_k\in\mathbb{R}^{n}$, $y_k\in\mathbb{R}$, and $u_k\in\mathbb{R}$ are the agent $k$'s state, output and input, respectively, \textcolor{black}{and $u_0=0$}. The function $\theta_k(t)$, called sensor sensitivity, is the ratio of the $k$-th sensor's output to its input, and the nonlinear terms $F_k =[f_{k,1} ,\cdots,f_{k,n}]^{\top}\in\mathbb{R}^n$ model the system uncertainties where $f_{k,i}$ is piecewise continuous in $t$ and smooth w.r.t. $x$. For system (\ref{model}), we need the following assumptions.
\begin{assumption}\label{ass-1}
	For $i=1,\cdots,n$, there exist constants $\varrho_{k,i}$ such that
	\begin{equation}\label{growth}
	|f_{k,i}(t,x_k)-f_{0,i}(t,x_0)|\leq \sum_{j=1}^{i}\varrho_{k,j} |x_{k,j}-x_{0,j}|.
	\end{equation}	
\end{assumption}
\begin{assumption}\label{ass-2}
	\textcolor{black}{For $k=0,1,\cdots,N$,} the sensor sensitivity \textcolor{black}{$\theta_k(t)$} is an unknown continuous function satisfying $\textcolor{black}{\theta_k(t)}\in [1-\textcolor{black}{\Delta_{\theta_k}}, 1 +\textcolor{black}{\Delta_{\theta_k}}],$ where $0 < \textcolor{black}{\Delta_{\theta_k}}< 1$ denotes the sensitivity error since it allows the sensor sensitivity has a  fluctuation of $\textcolor{black}{\pm \Delta_{\theta_k}}$.
\end{assumption}

%[{[\citen{2018-chen-tac,2020-auto-liwuquan}]}]
%写异构多智能体 %添加灵敏度未知的描述
\begin{remark}
It is evident that the multi-agent system (\ref{model}) can be heterogeneous since both the nonlinearity of each agent and the sensor used to measure the output of each agent can be different. 
		%Addressing the prescribed-time control for (\ref{model}) in cases where the dimensions of $\mathbf{A}$ and $B$ are different poses an intriguing yet formidable challenge that will be addressed in our forthcoming research.
\end{remark}

%For example, the robotic manipulator in \cite{2015-TFS-model} has the following form, for $k=0,1$:
%\begin{equation}\label{2}
%\dot{x}_{k,1}=x_{k,2},~\dot{x}_{k,2}=u_k+f_{k,2}(x_{k,1},x_{k,2}),
%\end{equation}
%where  $f_{k,2}=-{B_k}x_{k,2}/{J_k}-m_kgh_k\sin(x_{k,1}/J_k)$, and $B_k$, $J_k$, $m_k$, and $h_k$ are some physical parameters. It can be verified that Assumption \ref{ass-1} holds for system (\ref{2}) since
%\begin{equation}\label{4}
%\begin{aligned}
%|f_{1,2}-f_{0,2}|=&\left|{B_0}x_{0,2}/{J_0}+m_0gh_0\sin\left({x_{0,1}}/{J_0}\right)\right.\\&\left.-{B_1}x_{1,2}/{J_1}+m_1gh_1\sin\left( {x_{1,1}}/{J_1}\right)\right|\\
%\leq& ~ \varrho_{1,1}|x_{1,1}-x_{0,1}|+\varrho_{1,2}|x_{1,2}-x_{0,2}|,
%\end{aligned}
%\end{equation}
%where the growth rates $\varrho_{1,1}=\max\{m_0gh_0/J_0,m_1gh_1/J_1\}$ and $\varrho_{1,2}=\max\{B_0/J_0,B_1/J_1\}$. %Note that the growth rates might be unknown, which is a more relaxed assumption compared to those made in literature that achieve asymptotic stability/consensus, as they require require \textit{a priori} knowledge of the growth rate (see, for instance, \cite{2015-Auto-Zhang,2018-chen-tac} and the reference therein). 

We use the graph to describe the information exchange between agents and the leader. The agents indexed by  $1, \cdots,N$ are followers, while the agent indexed by $0$ is the leader. The communication topology among $N$ followers is represented by a weighted undirected graph $\mathcal{G}=\{\mathcal{V},\mathcal{E},\mathcal{A}\}$, where $\mathcal{V}=\{1,\cdots,N\}$ denotes the agent in the group, the set of edges $\mathcal{E}=\{(i,j)\in \mathcal{V}\times \mathcal{V}\}$ represents neighboring relations among  agents, and $\mathcal{A}=(a_{i,j})\in\mathbb{R}^{N\times N}$ is the weighted adjacency matrix of graph $\mathcal{G}$ as \textcolor{black}{$a_{i,j}=1$ if $(i,j)\in\mathcal{E}$ and $a_{i,j}=0$ otherwise.} The Laplacian matrix of $\mathcal{G}$ is given by $\mathcal{L}_{\mathcal{G}}=\mathcal{D}-\mathcal{A}$ where $\mathcal{D}$ is the degree matrix with $i$th diagonal elements $d_i=\sum_{j=1,j\neq i}^{N}a_{i,j}$ and $0$ elsewhere. Let $\bar{\mathcal{G}}$ be a graph generated by graph $\mathcal{G}$ and the leader. {Assuming that the graph ${\mathcal{G}}$ is connected,} and there is at least one directed edge from a follower to the leader, with no path leading from followers to  the leader. Furthermore, it is either known to all followers what the input of the leader is \cite{2015-Auto-Zhang}, or the input of the leader is zero\cite{2018-Auto-Wang}. The connection weight matrix is given by $\mathcal{B}=\operatorname{diag}\{\mathsf{b}_1,\cdots,\mathsf{b}_N\}$ as $\mathsf{b}_i=1$ if agent $i$ is a neighbor of the leader and $\mathsf{b}_i=0$ otherwise. The Laplacian matrix   $\mathcal{L}_{\bar{\mathcal{G}}}= \mathcal{L}_{\mathcal{G}}+\mathcal{B}$ associated with $\bar{\mathcal{G}}$ is positive definite\cite{2012-IJCAS-Wang}.

%In this network, we consider the problem of designing $u_i$, $i = 1, \cdots, N$ using local information to render all $N$ agents to follow the leader within a prescribed time. We make this precise by the following  definition.

\begin{definition}\label{definition-1}
	\textcolor{black}{The prescribed-time globally uniformly  consensus  problem of the leader-following MAS (\ref{model}) is said to be solved if, for  $k\in\{1,\cdots,N\}$,  there exist a class $\mathcal{KL}$ function $\beta$ and a time-varying function $\mu:[0,T)\rightarrow[0,\infty)$ such that $\mu(t)$ tends to infinity as $t$ goes to $T$, and $\forall x_k,x_0\in\mathbb{R}^n$
		\begin{equation}
		|x_k(t)-x_0(t)|\leq \beta\left(|x_k(0)-x_0(0)|,\mu(t)\right),~\forall t\in [0,T),
		\end{equation}  
		where  $T>0$ is an arbitrary time-independent constant.}
\end{definition}

%\begin{definition}\label{definition-2}
%The leader-following  MAS (\ref{model}) is said to be practical prescribed-time consistent if it is stable in the usual sense and $\lim_{t\rightarrow t_f}\|x_k(t)-x_0(t)\|\leq\Delta, ~\lim_{t\rightarrow \infty}\|x_k(t)-x_0(t)\|=0,~\forall x_k(0)\in\mathbb{R}^n$ with $\Delta$ being a small constant.
%\end{definition}
%
%The control objectives in this paper can then be stated as follows.
%
%\textit{Objective 1:} Let $t_f$ and $\delta$ be positive constants. Design distributed protocols $u_k(t)$ on $t\in[0,t_f+\delta)$ for system (\ref{model}), which is bounded for any given initial condition, such that the leader-following prescribed-time consensus  can be achieved in the sense of Definition \ref{definition-1}.
%
%\textit{Objective 2:} Let $t_f$ and $\delta$ be positive constants. Design distributed protocols $u_k(t)$ on $t\in[0,\infty)$ for system (\ref{model}), which is bounded for any given initial condition, such that the leader-following practical prescribed-time consensus can be achieved in the sense of Definition \ref{definition-2}.

\section{Preliminary result: \textcolor{black}{State-feedback without unknown measurement sensitivity}}\label{sec-state-feedback}

The matrix pencil based design framework was originally proposed in \cite{2020-NYU-IFAC,2020-CDC-matrix-pencil,2022-scl-matrix-pencil,2023-tac-matrix-pencil}, \textcolor{black}{and the robust and decentralized formulations of such procedure in the presence of additive noise and interactions were studied in \cite{2023-tac-ye-matrix-pencil} and \cite{2023-auto-ye-matrix-pencil}, respectively. In this paper, we propose for the first time a class of matrix pencil formulation robust to multiplicative noise for distributed prescribed-time consensus protocol design.} 
For ease of exposition, the matrix pencil based approach is first tested under state-feedback design. 

To begin with, we introduce the time-axis transformation, inspired by \cite{2017-DSMC-Kan}, to facilitate the convergence analysis of the prescribed-time consensus. Define, $\forall t\in[0,T) $, 
\begin{equation}\label{time-transformation}
t=T(1-e^{-\tau}) ~\Leftrightarrow~\tau=\ln\left(\frac{T}{T-t}\right).
\end{equation}   
Then, it can be verified that when $t=0$, $\tau=0$, and when  $t=T$, $\tau=\infty$, and
\begin{equation}\label{6}
\frac{d\tau}{dt}=\frac{1}{T}e^{\tau}\triangleq \alpha(\tau) .
\end{equation}
Inserting the expression of $\tau$ into (\ref{6}), it is seen that the equivalent form of $\alpha(\tau)$ on the $t$-axis is\footnote{Given a general signal $q$ of appropriate dimension, $q(\tau)=q(t)$ holds since $q(\tau)$ and $q(t)$ refer to the value	of the same signal  at the $\tau$-axis and $t$-axis, respectively.} 
\begin{equation}\label{alpha-t}
\alpha(t)=\frac{1}{T-t}  .
\end{equation} 
With (\ref{6}), the dynamics of  $\dot{x}=f(x,u,t)$ is converted into 
\begin{equation}\label{x-tau}
\frac{dx}{d\tau}=\frac{dt}{d\tau}f(x,u,\tau)=\frac{1}{\alpha(\tau)}f(x,u,\tau).
\end{equation}
Consequently, the prescribed-time control of $\dot{x}=f(x,u,t)$ can be equivalently converted into the  asymptotic control of (\ref{x-tau}), as formally stated in the following Proposition with its proof provided in the Appendix A.

%Specifically, as shown in (\ref{alpha0-t}),  we choose a function  that monotonously increases on $[0,t_f)$, and add a saturation with a magnitude of $1/\delta$ to $\mu$ at $t=t_f$. The analysis in Section \ref{sec-output} shows that there is a quantitative relationship between the magnitude of saturation and the selection of controller parameters.}

\begin{proposition}\label{proposition-1}
	\textcolor{black}{If the coordinate transformation (\ref{eta_k}) is applied,}
	\begin{equation}\label{eta_k}
	\textcolor{black}{\eta_{k}=\operatorname{diag} \{r^{n},r^{n-1},\cdots,r \}(x_{k}-x_{0}) :=\mathbf{\Lambda}_r(x_{k}-x_{0}),} 
	\end{equation}
	\textcolor{black}{and the transformed dynamic system (\ref{model-state}) is asymptotically stabilized,\footnote{\textcolor{black}{For notational convenience, we drop the arguments of functions whenever no confusion will occur.}}}
	\begin{equation} \label{model-state} 
	\textcolor{black}{\begin{aligned}
		\frac{d\eta_{k}}{d\tau} =b\left(\mathbf{A}\eta_{k}+B u_k +\Phi_k\right)+\mathbf{D}_c\eta_k,
		\end{aligned}}
	\end{equation} 
	\textcolor{black}{where $r=be^{\tau}/T $ with   $b \geq T$, $\mathbf{D}_c=\operatorname{diag}\{n,n-1,\cdots,1\}$, $\Phi_k=[\phi_{k,1},\cdots,\phi_{k,n}]^{\top}$ with $\phi_{k,i}=r^{n-i}(f_{k,i}-f_{0,i})$, namely, there exists a class $\mathcal{KL}$ function $\beta$ such that, for $\eta=[\eta_1^{\top},\cdots,\eta_N^{\top}]^{\top}\in\mathbb{R}^{nN}$,}
	\begin{equation}
	\textcolor{black}{|\eta (\tau)|\leq \beta\left(|\eta (0)|,\tau\right),~\forall \tau \in[0,\infty),}
	\end{equation} 
	\textcolor{black}{then the prescribed-time consensus of the original leader-following MAS (\ref{model}) is achieved in the sense of Definition 1.}
\end{proposition}

Now, we are in the position to state our preliminary result.
\begin{theorem}\label{theorem-2}
	Suppose that the graph \textcolor{black}{${\mathcal{G}}$} is connected and Assumption \ref{ass-1} is satisfied with unknown growth rates $\varrho_{k,j}$. A proportional coefficient $b$ can be found such that system (\ref{model-state}) is asymptotically stabilized by a distributed protocol:
	\begin{equation}\label{u_k_1}
	\textcolor{black}{u_k =-K\mathbf{\Lambda}_r\Big( \sum_{j=1}^{N}a_{k,j}( {x}_k- {x}_j)+\mathsf{b}_k( {x}_k- {x}_0) \Big) ,}
	\end{equation}
	where  $K$ is chosen such that $\mathbf{A}_c=\mathbf{I}_{N}\otimes \mathbf{A}-\mathcal{L}_{\bar{\mathcal{G}}}\otimes(BK)$ is Hurwitz.  	
\end{theorem}
\textit{Proof:} Let $\mathbf{D} =\mathbf{I}_N\otimes\mathbf{D}_c\textcolor{black}{\in\mathbb{R}^{nN\times nN}}$ and $\Phi=[\Phi_1^{\top},\cdots,\Phi_N^{\top}]^{\top}\in\mathbb{R}^{nN}.$ Since the equivalent form  of protocol (\ref{u_k_1}) is $\textcolor{black}{u_k=  -(\Gamma_k\otimes K)\eta} $ where $\Gamma_k$ is the $k$th row of $\mathcal{L}_{\bar{\mathcal{G}}}$,  then system (\ref{model-state}) can be written in a compact form as follows
\begin{equation}\label{d-eta}
\frac{d\eta}{d\tau}=b \left(\mathbf{A}_{c} \eta+\Phi\right) +   \mathbf{D} \eta.
\end{equation} 
\textcolor{black}{Choose a Lyapunov-like  function candidate as}
\begin{equation}\label{12}
V_c(\tau) =\eta^{\top}\mathbf{P}_c\eta,
\end{equation}
where $\mathbf{P}_c\in\mathbb{R}^{nN\times nN}$ is symmetric positive-definite (SPD), then the derivative of $V_c$ along the trajectory of (\ref{d-eta}) is
\begin{equation}\label{dV-c}
\begin{aligned}
\frac{dV_c}{d\tau}&=b \eta^{\top}\left(\mathbf{P}_c\mathbf{A}_c+\mathbf{A}_c^{\top}\mathbf{P}_c\right)\eta+2b\eta^{\top}\mathbf{P}_c\Phi \\
&~~~+  \eta^{\top}\left(\mathbf{P}_c\mathbf{D}+\mathbf{D}\mathbf{P}_c\right) \eta.
\end{aligned}
\end{equation}
Now the design purpose is to find parameters $b$, $\kappa_0$, and $\kappa_1$ such that (\ref{dV-c}) satisfies the following form:
\begin{equation}\label{dV-2}
\begin{aligned}
\frac{dV_c}{d\tau}\leq-\left(\kappa_0-\frac{\kappa_1}{\alpha(\tau)}\right)V_c.
\end{aligned}
\end{equation}
\textcolor{black}{To this point, the design procedure can be divided into  two steps.}

\textcolor{black}{\textbf{\textit{Step 1:}} $b$ is selected to ensure that the following inequality is satisfied with $\kappa_{0}$ being an arbitrary positive constant:}
\begin{equation} \label{14}
\textcolor{black}{\begin{aligned}
	b \eta^{\top}\left(\mathbf{P}_c\mathbf{A}_c+\mathbf{A}_c^{\top}\mathbf{P}_c\right)\eta   &\\
	+  \eta^{\top}\left(\mathbf{P}_c\mathbf{D}+\mathbf{D}\mathbf{P}_c\right) \eta +\kappa_0\eta^{\top}\mathbf{P}_c\eta&\leq 0.
	\end{aligned}}
\end{equation} 
Notice that $\mathbf{A}_c$ is Hurwitz and $\mathbf{D}$ is SPD, hence  a SPD matrix $\mathbf{P}_c$ can be found such that
\begin{equation}\label{17}
\begin{aligned}
\mathbf{P}_c\mathbf{A}_c+ \mathbf{A}_c^{\top}\mathbf{P}_c& \leq - \mathbf{I}_{nN}, \\
\mathbf{P}_c\mathbf{D}+\mathbf{D} \mathbf{P}_c  &\geq \mathbf{I}_{nN}. 
\end{aligned}
\end{equation} 
\textcolor{black}{Therefore, one can conclude that $\mathbf{P}_c\mathbf{D}+\mathbf{D} \mathbf{P}_c+\kappa_0\mathbf{P}_c$ is SPD and $\mathbf{P}_c\mathbf{A}_c+\mathbf{A}_c^{\top}\mathbf{P}_c$ is symmetric negative-definite (SND). Then, pursuant to Lemma \ref{lemma-matrix}, we can pre-give a $\kappa_0$ and choose}
\begin{equation}\label{26}
\textcolor{black}{b=\max\left\{\sigma\left(\mathbf{Q}_{1},-\mathbf{Q}_{2}\right),T\right\},}
\end{equation}
where $\mathbf{Q}_1=\mathbf{P}_c\mathbf{D}+\mathbf{D} \mathbf{P}_c+\kappa_0\mathbf{P}_c$ and $\mathbf{Q}_2=\mathbf{P}_c\mathbf{A}_c+\mathbf{A}_c^{\top}\mathbf{P}_c$, such that (\ref{14}) always holds. \textcolor{black}{Since the calculation of $b$ only requires local information $\mathbf{A}$, $B$ and $K$, and does not require information about neighbor agents and the leader, the resulting controller is parameterly decentralized.}

\textcolor{black}{\textbf{\textit{Step 2:}} Here we aim to verify that the following inequality  holds for some positive constant $\kappa_1$,}
\begin{equation}\label{17-1} 
\textcolor{black}{\begin{aligned}
	2b\eta^{\top}\mathbf{P}_c\Phi  \leq \frac{\kappa_1}{\alpha(\tau)} V.
	\end{aligned}}
\end{equation} 
Under Assumption \ref{ass-1} and the fact $r\geq 1$, it can be obtained that
\begin{equation}
|\Phi|_e  \leq_e  \frac{1}{r}\operatorname{diag}\{\mathbf{A}_{\varrho_1},\cdots,\mathbf{A}_{\varrho_N}\}|\eta|_e:=\frac{1}{r}\mathbf{A}_{\varrho}|\eta|_e,
\end{equation}
where
\begin{equation}\label{Arhok}
\begin{array}{ll}
\mathbf{A}_{\varrho_k} =
\left[\begin{array}{ccccc} 
\varrho_{k,1} & 0 &\cdots & 0 \\
\varrho_{k,1}  & \varrho_{k,2}  & \cdots &0 \\
\vdots & \vdots & \ddots & \vdots \\
\varrho_{k,1} &\varrho_{k,2} &\cdots&\varrho_{k,n} 
\end{array}\right]\in\mathbb{R}^{n\times n}.
\end{array}
\end{equation} 
Therefore, \textcolor{black}{by recalling $r=b\alpha(\tau)$,} we get 
\begin{equation}\label{16}
\textcolor{black}{2b\eta^{\top}\mathbf{P}_c\Phi \leq \frac{1 }{\alpha(\tau)}|\eta|_e^{\top}(\mathbf{A}_\varrho^{\top}|\mathbf{P}_c|_e+|\mathbf{P}_c|_e\mathbf{A}_\varrho)|\eta|_e .}
\end{equation} 
\textcolor{black}{Let $\bar{\mathbf{P}}_c=\operatorname{diag}\{\mathbf{P}_c\}$   be a positive definite diagonal matrix. From Lemma \ref{lemma-matrix},  there exist two positive constants $\kappa_1$ and $\kappa_2$, such that,}
\begin{equation}\label{Matrix-P}
\textcolor{black}{\begin{aligned}
	-\kappa_2 \bar{\mathbf{P}}_c 
	+  	   (\mathbf{A}_\varrho^{\top}|\mathbf{P}_c|_e+|\mathbf{P}_c|_e\mathbf{A}_\varrho)&\leq 0,\\
	-\kappa_1  {\mathbf{P}}_c +
	\kappa_2\bar{\mathbf{P}}_c & \leq 0 . 
	\end{aligned}} 
\end{equation}
\textcolor{black}{Specifically, the suitable $\kappa_1$ and $\kappa_2$ are given by}
\begin{equation}\label{kappa}
\textcolor{black}{\begin{aligned} 
	\kappa_1 &=\max\{\sigma(\kappa_2\bar{\mathbf{P}}_{c},\mathbf{P}_{c})\},\\ \kappa_2 &=\max\{\sigma( \mathbf{A}_\varrho^{\top}|\mathbf{P}_c|_e+|\mathbf{P}_c|_e\mathbf{A}_\varrho, \bar{\mathbf{P}}_{c})\}.
	\end{aligned}\hspace{-1em}}
\end{equation} 
Now we can upper-bound the right-hand side of (\ref{16}), in view of (\ref{Matrix-P}) and the fact $ \eta^{\top}\bar{\mathbf{P}}_c \eta=| \eta|_e^{\top}\bar{\mathbf{P}}_c| \eta|_e$,  as 
 \begin{align}
\nonumber	\frac{1 }{\alpha(\tau)}|\eta|_e^{\top}(\mathbf{A}_\varrho^{\top}|\mathbf{P}_c|_e+|\mathbf{P}_c|_e\mathbf{A}_\varrho)|\eta|_e&\leq  \frac{\kappa_2}{\alpha(\tau)} |\eta|_e^{\top}\bar{\mathbf{P}}_c|\eta|_e,\\
	\frac{\kappa_2}{\alpha(\tau)} |\eta|_e^{\top}\bar{\mathbf{P}}_c|\eta|_e=\frac{\kappa_2}{\alpha(\tau)}  \eta ^{\top} \bar{\mathbf{P}}_c \eta &\leq \frac{\kappa_1}{\alpha(\tau)}  \eta ^{\top} {\mathbf{P}}_c \eta .\label{kappa-2}
	\end{align}  
Recalling (\ref{12}), (\ref{16}), and (\ref{kappa-2}),  {we know that} (\ref{17-1}) holds. Combining (\ref{dV-c}), (\ref{14}), and (\ref{17-1}), it is verified that (\ref{dV-2}) holds.  It follows from Lemma \ref{lemma-2} that
\begin{equation}\label{vv}
\textcolor{black}{V_c(\tau)\leq e^{-\kappa \tau}MV_c(0),}
\end{equation}
\textcolor{black}{where $\kappa=\kappa_0-\kappa_1/\alpha(\tau_1)>0$ and $M=e^{T\kappa_1\tau_1}$ is a finite number with $0<\tau_1<\infty$. Insert $\tau$ defined in (\ref{time-transformation}) into (\ref{vv}) yields}
\begin{equation}\label{54-1}
\textcolor{black}{V_c(t)\leq \left(\frac{\textcolor{black}{T}-t}{\textcolor{black}{T}}\right)^{\kappa}MV_c(0) ,~t\in[0,\textcolor{black}{T}).}
\end{equation} 
Therefore, it holds that $\lim_{t\rightarrow \textcolor{black}{T}}V_c(t)=0$ and $V_c(t)\in\mathcal{L}_\infty[0,\textcolor{black}{T})$. Additionally, it follows from (\ref{12}) that\footnote{\textcolor{black}{This estimate does not lead to conservative selection of control parameters since it  is only involved in the stability analysis, that is, only the  transient performance  of system state and   the input are conservatively estimated.}}
\begin{equation}\label{54}
\textcolor{black}{|\eta(t)| \leq \sqrt{\frac{{V_c(t)}}{ \lambda_{\min}(\mathbf{P}_c) }} .}
\end{equation}  
With $\mathbf{\Lambda}_r\hspace{-0.2em}=\hspace{-0.2em}\operatorname{diag}\{r^{n},r^{n-1},\cdots,r\}$ and the fact $r\hspace{-0.2em}\geq \hspace{-0.2em}1$, it holds that
\begin{equation}\label{55-1}
\textcolor{black}{\begin{aligned}
	|x_k(t)-x_0(t)|\leq \frac{|\eta(t)|}{r}  =\frac{(\textcolor{black}{T}-t)^{\frac{\kappa+2}{2}}}{bT^{\kappa/2}}\sqrt{\frac{{MV_c(0)}}{ \lambda_{\min}(\mathbf{P}_c) }} .
	\end{aligned}}
\end{equation}
Hence, we obtain $x_k-x_0 \in\mathcal{L}_\infty[0,\textcolor{black}{T})$ and $\lim_{t\rightarrow T}|x_k-x_0|=0$.  By recalling (\ref{u_k_1}), we know that the designed control inputs satisfy $u_k \in\mathcal{L}_\infty[0,\textcolor{black}{T})$.   
The proof is completed. $\hfill\blacksquare$

\begin{remark}
	Note that calculating $\kappa_2$ requires \textit{a prior} knowledge of the growth rate $\varrho_{k,i}$ that is contained in $\mathbf{A}_{\varrho}$, and calculating $\kappa_1$ requires \textit{a prior} knowledge of $\kappa_2$. \textcolor{black}{Since we use $\kappa_1$ and $\kappa_2$ only for stability analysis rather than controller design, \textit{a prior} knowledge of $\varrho_{k,i}$ is not required, which makes the proposed algorithm more general compared to the algorithms presented in \cite{2018-chen-tac} and \cite{2015-Auto-Zhang}.} 
\end{remark}

%\begin{remark} 
%	Protocol (\ref{u_k}) can be equivalently written as 
%	\begin{equation*} 
%	u_k =-K\mathbf{\Lambda}_r\Big(\sum_{j=1}^{N}a_{k,j}( {x}_k- {x}_j)+b_k( {x}_k- {x}_0)\Big)+u_0,
%	\end{equation*}
%	which is constructed by the local neighborhood errors represented by $\sum_{j=1}^{N}a_{k,j}({x}_k-{x}_j)+b_k({x}_k-{x}_0)$. It is seen that 
%	such a protocol is fundamentally different from the finite- and fixed-time protocols (see, for instance, \cite{2018-IJC-Ning-finite=fix}) as it is non-autonomous, and the feedback signals are smooth rather than non-smooth. If the input of the leader is set to $u_0\equiv0$, the proposed scheme becomes full distributed protocol, as considered in \cite{2018-Auto-Wang}. %It is worth mentioning that the controller is essentially time-varying linear feedback and does not involve any recursive computation, which simplifies the analysis of the boundedness of the control input and reduces the computational complexity of the algorithm.
%\end{remark}	

\begin{remark} 
	Note that the scaling order of the coordinate transformations (\ref{eta_k}) in our scheme has been subtly changed compared to that of the state-of-the-art works \cite{2020-NYU-dynamic-high-gain,2021-NYU-ACSP,2021-RNC-NYU}. One of the benefits of this is that it technically allows the design of a static rather than dynamic feedback control law thus directly avoiding the computational problem of integrating over an infinite number. Another advantage is that the prescribed-time convergence property of the tracking error (i.e., $x_{k,i}-x_{0,i}$) can be obtained directly from the asymptotic convergence property of $\eta_k$. Specifically,  the coordinate transformation might be defined as $x_{k,i}-x_{0,i}=r^{i}\eta_{k,i}$ if we adopt the control scheme proposed in \cite{2020-NYU-dynamic-high-gain,2021-NYU-ACSP,2021-RNC-NYU}. It can be seen that further proof that $\eta_{k,i}$ decays faster than the growth rate of $r^i$ is needed to obtain the conclusion that the tracking error converges to zero.
\end{remark}

\section{Main results: \textcolor{black}{Output-feedback with unknown measurement sensitivity}}\label{sec-output}

\subsection{Prescribed-time exact consensus tracking}

In this section, we show that  using only inaccurate output measurements $y_k=\textcolor{black}{\theta_k(t)}x_{k,1}$, the prescribed-time consensus can be achieved. \textcolor{black}{Using the time-varying parameter $r$ and the diagonal matrix $\mathbf{\Lambda}_r$ defined in (\ref{eta_k}),   the prescribed-time distributed observer can be constructed as,} 
\begin{equation}\label{observer}
\textcolor{black}{\begin{aligned} 
	\dot{\hat{x}}_{k}&=\mathbf{A}\hat{x}_{k}+Bu_k+r^{n+1}\mathbf{\Lambda}_r^{-1}G(y_k-\hat{x}_{k,1}) ,
	\end{aligned}}
\end{equation}
where $G=[g_1,\cdots,g_n]$. Correspondingly, the prescribed-time consensus tracking problem can be transformed into an asymptotic stabilization problem of the neighborhood error's dynamics, as formally stated in the following proposition. 
\begin{proposition}\label{proposition-2}
	\textcolor{black}{If the coordinate transformation (\ref{29}) is applied,}
	\begin{equation}\label{29}
	\textcolor{black}{\begin{aligned}
		\hat\eta_{k} &=\mathbf{\Lambda}_r(\hat x_{k}-\hat x_{0}),~\\
		\varepsilon_k &=\mathbf{\Lambda}_r(x_{k}-x_{0})-\hat\eta_k,
		\end{aligned}}
	\end{equation}
	\textcolor{black}{and the transformed dynamic system (\ref{30}) is asymptotically stabilized,}  
	\begin{equation}\label{30}
\begin{aligned}
		\frac{d{\hat{\eta}_k}}{d\tau}&=b\left(\mathbf{A}\hat{\eta}_k+\textcolor{black}{B u_k} \right)+\mathbf{D}_c\hat{\eta}_k\\
		&~~~ +bG^{\top}\theta_k \varepsilon_{k,1} +bG^{\top}(\theta_k-1)\hat{\eta}_{k,1},\\
		\frac{d\varepsilon_k}{d\tau}&=b\left( (\mathbf{A}-G^{\top}C)\varepsilon_k+\Phi_k\right)+ \mathbf{D}_c\varepsilon_k,
		\end{aligned}
	\end{equation} 
	\textcolor{black}{namely, there exists a class $\mathcal{KL}$ function $\beta$ such that, for $\hat{\eta}=[\hat\eta_1^{\top},\cdots,\hat\eta_N^{\top}]^{\top}\in\mathbb{R}^{nN}$ and $\varepsilon=[\varepsilon_1^{\top},\cdots,\varepsilon_N^{\top}]^{\top}\in\mathbb{R}^{nN}$,} 
	\begin{equation}
	\textcolor{black}{|\hat\eta (\tau)+\varepsilon(\tau) |\leq \beta\left(|\hat\eta (0)+\varepsilon(0)|,\tau\right),~\forall \tau\in[0,\infty),}
	\end{equation}  
	\textcolor{black}{where $\mathbf{A}$, $B$, $C$, $\mathbf{D}_c$, $\Phi_k$, and $\tau$ follow the definitions in the previous Section, and $G$ is chosen such that $\mathbf{A}_0=\mathbf{I}_N\otimes(\mathbf{A}-G^{\top}C) $ is Hurwitz,	then the prescribed-time consensus of the original leader-following MAS (\ref{model}) is achieved in the sense of Definition 1.}
\end{proposition}

\textit{Proof:} \textcolor{black}{This proposition can be proven by a direct extension of the proof of Proposition \ref{proposition-1} and the details are omitted for brevity.}

\textcolor{black}{With Proposition \ref{proposition-2}, we can establish the following theorem.}
\begin{theorem}\label{theorem-3}
	Suppose that the graph \textcolor{black}{${\mathcal{G}}$} is connected and Assumptions \ref{ass-1} and  \ref{ass-2} are satisfied with unknown growth rates $\varrho_{k,j}$. \textcolor{black}{If the maximum value of the allowable sensitivity error satisfies} 
	\begin{equation}\label{allowable}
	\textcolor{black}{0<\max \{\Delta_{\theta_1},\cdots,\Delta_{\theta_N}\}\leq  \frac{1}{ |\mathbf{P}_c\mathbf{A}_g|},}
	\end{equation}
	where $\mathbf{A}_g$ is defined in (\ref{Ag}), then a  proportional coefficient $b$ can be found such that system (\ref{model-state}) is asymptotically stabilized by the distributed observer (\ref{observer}) and the following output-feedback distributed protocol
	\begin{equation}\label{32}
	\begin{aligned}
	\textcolor{black}{u_k =-K\mathbf{\Lambda}_r\Big( \sum_{j=1}^{N}a_{k,j}(\hat{x}_k-\hat{x}_j)+\mathsf{b}_k( \hat{x}_k- \hat{x}_0) ,}
	\end{aligned}
	\end{equation}
	where $K$ is chosen such that $\mathbf{A}_c= \mathbf{I}_{N}\otimes \mathbf{A}-\mathcal{L}_{\bar{\mathcal{G}}}\otimes (BK)$ is Hurwitz. 	
\end{theorem}

\textit{Proof:}  Since the protocol (\ref{32}) can be rewritten as $\textcolor{black}{u_k=-(\Gamma_k\otimes K)\hat{\eta}} $ where $\hat{\eta}=[\hat{\eta}_1^{\top},\cdots,\hat{\eta}_N^{\top}]^{\top}\in\mathbb{R}^{nN}$ and $\Gamma_k$ is the $k$th row of $\mathcal{L}_{\bar{\mathcal{G}}}$, then the following compact form is obtained, in view of (\ref{30}), 
\begin{equation}\label{33}
\begin{aligned}
\frac{d\hat\eta}{d\tau}&=b\left(\mathbf{A}_c\hat\eta+G^{\top}\otimes \left(\textcolor{black}{\mathbf{I}_\theta}\Psi_{\varepsilon}+\textcolor{black}{(\mathbf{I}_\theta-\mathbf{I}_N)}\Psi_{\hat\eta}\right)\right) + \mathbf{D} \hat\eta , \\
\frac{d\varepsilon}{d\tau}&=b\left(\mathbf{A}_0\varepsilon+\Phi\right) + \mathbf{D} \varepsilon ,
\end{aligned}
\end{equation} 
where  $\mathbf{D}$ and $\Phi$ are defined in (\ref{d-eta}), $
\textcolor{black}{\mathbf{I}_{\theta}} \textcolor{black}{=\operatorname{diag}\{\theta_1,\cdots,\theta_N\} },$ $ 
\Psi_{\varepsilon} =[\varepsilon_{1,1},\cdots,\varepsilon_{N,1}]^{\top} , $ and $  \Psi_{\hat\eta} =[\hat\eta_{1,1},\cdots,\hat\eta_{N,1}]^{\top} . $

Choose a \textcolor{black}{Lyapunov-like} function defined on $[0,\infty)$ as follows,
\begin{equation}\label{35}
V(\tau)=\hat\eta^{\top}\mathbf{P}_c\hat\eta+c\varepsilon^{\top}\mathbf{P}_0\varepsilon,
\end{equation}
where $\mathbf{P}_c $ and $\mathbf{P}_0  $ are SPD, and $c$ is a positive constant to be designed later in (\ref{c}). Calculating the derivative of $V$ along the trajectories of (\ref{33}) yields 
\begin{align}
\nonumber\frac{dV}{d\tau}= &~b  \hat\eta^{\top}\left(\mathbf{P}_c\mathbf{A}_c+\mathbf{A}_c^{\top}\mathbf{P}_c\right)\hat\eta+\hat\eta^{\top}\left(\mathbf{P}_c\mathbf{D}+\mathbf{D}\mathbf{P}_c\right)\hat\eta\\
\nonumber&+b  c\varepsilon^{\top}\left(\mathbf{P}_0\mathbf{A}_0+\mathbf{A}_0^{\top}\mathbf{P}_0\right)\varepsilon+c\varepsilon^{\top}\hspace{-0.05cm}(\mathbf{P}_0\mathbf{D}+\mathbf{D}\mathbf{P}_0)\varepsilon \\
\nonumber&+2bc\varepsilon^{\top}\mathbf{P}_0\Phi+ 2b \hat\eta^{\top}\mathbf{P}_c\left(G^{\top}\otimes(\textcolor{black}{\mathbf{I}_{\theta}}\Psi_{\varepsilon})\right) \\
&+ 2b\hat\eta^{\top}\mathbf{P}_c\left(G^{\top}\hspace{-0.1cm}\otimes\left(\textcolor{black}{(\mathbf{I}_\theta-\mathbf{I}_N)}\Psi_{\hat\eta}\right)\right).\label{dV}
\end{align} 	 
To deal with the last two uncertain terms in  (\ref{dV}), we define
\begin{equation}\label{Ag}
\mathbf{A}_g=\operatorname{diag}\{\mathbf{A}_{g,1},\cdots,\mathbf{A}_{g,N}\},
\end{equation} 
where each partitioned matrix $\mathbf{A}_{g,k}$ has the same structure, i.e.,  $\mathbf{A}_{g,k}=[G^{\top},\mathbf{0},\cdots,\mathbf{0}]\in\mathbb{R}^{n\times n}$, $k=1,\cdots,N$. Then $G^{\top}\otimes(\textcolor{black}{\mathbf{I}_{\theta}}\Psi_{\varepsilon})$ and $G^{\top}\otimes(\textcolor{black}{(\mathbf{I}_\theta-\mathbf{I}_N)}\Psi_{\hat\eta})$ can be equivalently written as
\begin{equation}\label{23}
\begin{aligned}
G^{\top}\otimes(\textcolor{black}{\mathbf{I}_{\theta}}\Psi_{\varepsilon}) &=\mathbf{A}_g(\textcolor{black}{\mathbf{I}_n\otimes \mathbf{I}_\theta})\varepsilon,\\
G^{\top}\otimes\left(\textcolor{black}{(\mathbf{I}_\theta-\mathbf{I}_N)}\Psi_{\hat\eta}\right) &=\mathbf{A}_g\textcolor{black}{(\mathbf{I}_n\otimes(\mathbf{I}_\theta-\mathbf{I}_N))}\hat\eta.
\end{aligned}
\end{equation}
By Assumption \ref{ass-2},  it holds that \textcolor{black}{$\theta_k\leq 1+\Delta_{\theta_k}$} and $|\textcolor{black}{\theta_k-1|\leq\Delta_{\theta_k}}$. Therefore,
\begin{equation}\label{24}
\textcolor{black}{\begin{aligned}
	|\mathbf{I}_\theta|_e &\leq_e |\mathbf{I}_N+ \mathbf{I}_{\Delta_\theta} |_e, ~\\
	|\mathbf{I}_\theta-\mathbf{I}_N|_e  &\leq_e |\mathbf{I}_{\Delta_\theta}|_e, 
	\end{aligned}}
\end{equation}
where $\textcolor{black}{\mathbf{I}_{\Delta_\theta}} \textcolor{black}{=\operatorname{diag}\{\Delta_{\theta_1},\cdots,\Delta_{\theta_N}\}\in\mathbb{R}^{N\times N}}$. \textcolor{black}{With (\ref{allowable}), one can verify}
\begin{equation}\label{42-2}
\textcolor{black}{0<|\mathbf{I}_N\otimes \mathbf{I}_{\Delta_\theta}|\leq\max \{\Delta_{\theta_1},\cdots,\Delta_{\theta_N}\}\leq \frac{1}{|\mathbf{P}_c\mathbf{A}_g|}.}
\end{equation}
\textcolor{black}{Due to (\ref{23}), we can upper-bound the last term in (\ref{dV}) as} 
\begin{equation}\label{49-2}
\textcolor{black}{\begin{aligned}  2b\hat{\eta}^{\top}\mathbf{P}_c&\left(G^{\top}\otimes\left((\mathbf{I}_\theta-\mathbf{I}_N)\Psi_{\hat\eta}\right)\right)   \\
	\leq &~2b |\hat{\eta}| |\mathbf{P}_c\mathbf{A}_g\mathbf{I}_N\otimes \mathbf{I}_{\Delta_\theta} |    |\hat{\eta}|  
	\leq  2b \hat{\eta} ^{\top} \hat{\eta}.
	\end{aligned}}
\end{equation}
\textcolor{black}{Let $\mathbf{P}_\varepsilon=\mathbf{P}_c\mathbf{A}_g(\mathbf{I}_n\otimes (\mathbf{I}_N+ \mathbf{I}_{\Delta_\theta}))$. It follows by 
	(\ref{23}) and (\ref{24})  that,}
\begin{equation}\label{49-1}
\textcolor{black}{\begin{aligned}
	2b\hat{\eta}^{\top}\mathbf{P}_c\left(G^{\top}\otimes(\textcolor{black}{\mathbf{I}_{\theta}}\Psi_{\varepsilon})\right)  \leq   2  b|\hat{\eta}|_e^{\top}\textcolor{black}{|\mathbf{P}_\varepsilon|_e}|\varepsilon|_e , 
	\end{aligned}}
\end{equation} 
By \textcolor{black}{(\ref{49-2}) and (\ref{49-1})}, (\ref{dV}) reduces to  
\begin{align}
\nonumber \frac{dV}{d\tau}\leq&~ b  \hat\eta^{\top}\left(\mathbf{P}_c\mathbf{A}_c+\mathbf{A}_c^{\top}\mathbf{P}_c\right)\hat\eta+\hat\eta^{\top}\left(\mathbf{P}_c\mathbf{D}+\mathbf{D}\mathbf{P}_c\right)\hat\eta\\
\nonumber &+b  c\varepsilon^{\top}\hspace{-0.05cm}\left(\mathbf{P}_0\mathbf{A}_0+\mathbf{A}_0^{\top}\mathbf{P}_0\right)\varepsilon+c\varepsilon^{\top}\hspace{-0.05cm}(\mathbf{P}_0\mathbf{D}+\mathbf{D}\mathbf{P}_0)\varepsilon\\
&+2bc\varepsilon^{\top}\mathbf{P}_0\Phi + 2b \left( \hat{\eta}  ^{\top}  \hat{\eta} +  |\hat{\eta}|_e^{\top}\textcolor{black}{|\mathbf{P}_\varepsilon|_e}|\varepsilon|_e\right).\label{dV-new}
\end{align} 
Now the deign purpose is to find  parameters $b$ and $c$ such that (\ref{dV-new}) satisfies the following form:
\begin{equation}\label{dV-final}
\begin{aligned}
\frac{dV}{d\tau}\leq-\left(\kappa_a-\frac{\kappa_b}{\alpha(\tau)}\right)V,
\end{aligned}
\end{equation}
\textcolor{black}{where $\kappa_a$ and $\kappa_b$ are two design parameters.}
To this point, the design procedure can be divided into  \textcolor{black}{three steps.}

\textbf{\textit{Step 1:}} $c$ is picked to ensure that the following inequality is satisfied \textcolor{black}{with $c_1\in(0,1)$:}
\begin{equation}\label{42}
\begin{aligned}
\textcolor{black}{(1-c_1)} \left(\hat\eta^{\top} (\mathbf{P}_c\mathbf{A}_c+\mathbf{A}_c^{\top}\mathbf{P}_c)\hat\eta \right.&\\ \left. +c\varepsilon^{\top} (\mathbf{P}_0\mathbf{A}_0+\mathbf{A}_0^{\top}\mathbf{P}_0  )\varepsilon\right)  +2   |\hat{\eta}|_e^{\top}\textcolor{black}{|\mathbf{P}_\varepsilon|_e}|\varepsilon|_e &\leq 0.
\end{aligned}
\end{equation}
In order to convert (\ref{42}) into a matrix pencil structure, we need to find constants $\delta_{{A}_c}$ and $\delta_{{A}_0}$ such that following inequalities hold, 
\begin{equation}\label{34-C}
\begin{aligned}
&\textcolor{black}{\delta_{{A}_c}\hat\eta^{\top}\bar{\mathbf{A}}_c\hat\eta \geq \hat\eta^{\top} (\mathbf{P}_c\mathbf{A}_c+\mathbf{A}_c^{\top}\mathbf{P}_c )\hat\eta},\\
&\delta_{{A}_0}\varepsilon^{\top}\bar{\mathbf{A}}_0\varepsilon \geq \varepsilon^{\top} (\mathbf{P}_0\mathbf{A}_0+\mathbf{A}_0^{\top}\mathbf{P}_0 )\varepsilon, 
\end{aligned}
\end{equation} 
where $
\bar{\mathbf{A}}_c =\operatorname{diag}\{\mathbf{P}_c\mathbf{A}_c+\mathbf{A}_c^{\top}\mathbf{P}_c\} $ and 
$\bar{\mathbf{A}}_0 =\operatorname{diag}\{\mathbf{P}_0\mathbf{A}_0+\mathbf{A}_0^{\top}\mathbf{P}_0\}$ 
are diagonal matrices. Notice that $\mathbf{A}_c$ and $\mathbf{A}_0$ are Hurwitz, hence two SPD matrices $\mathbf{P}_c$ and $\mathbf{P}_0$ can be found such that the following  inequalities hold 
\begin{equation}\label{58}
\begin{aligned}
c_1(\mathbf{P}_c\mathbf{A}_c+\mathbf{A}_c^{\top}\mathbf{P}_c)  &\leq - {2} \mathbf{I}_{nN},~\\
\mathbf{P}_0\mathbf{A}_0+\mathbf{A}_0^{\top}\mathbf{P}_0  &\leq -\mathbf{I}_{nN}.
\end{aligned}
\end{equation}
It follows by, the fact $\bar{\mathbf{A}}_c$ and $\bar{\mathbf{A}}_0$ are  SND and Lemma \ref{lemma-matrix}, that (\ref{34-C}) holds for
\begin{equation}\label{59-A}
\begin{aligned}
&\delta_{{A}_c}= \min\{\sigma(\mathbf{P}_c\mathbf{A}_c+\mathbf{A}_c^{\top}\mathbf{P}_c,\bar{\mathbf{A}}_c)\},\\
&\delta_{{A}_0}= \min\{\sigma(\mathbf{P}_0\mathbf{A}_0+\mathbf{A}_0^{\top}\mathbf{P}_0,\bar{\mathbf{A}}_0)\}.\\
\end{aligned}
\end{equation} 
Since $\bar{\mathbf{A}}_c$ and $\bar{\mathbf{A}}_0$ are diagonal matrices, then we have $\hat\eta^{\top}\bar{\mathbf{A}}_c\hat\eta=|\hat\eta|_e^{\top}\bar{\mathbf{A}}_c|\hat\eta|_e$ and $\varepsilon^{\top}\bar{\mathbf{A}}_0\varepsilon=|\varepsilon|_e^{\top}\bar{\mathbf{A}}_0|\varepsilon|_e$. With (\ref{34-C}), we know that (\ref{42}) is a consequence of the following inequality:
\begin{equation}\label{42-A}
\begin{aligned}
\textcolor{black}{(1-c_1)}\left(\delta_{{A}_c}|\hat\eta|_e^{\top}\bar{\mathbf{A}}_c|\hat\eta|_e +c\delta_{{A}_0} |\varepsilon|_e^{\top}\bar{\mathbf{A}}_0|\varepsilon|_e\right)&\\ +2   |\hat{\eta}|_e^{\top}\textcolor{black}{|\mathbf{P}_\varepsilon|_e}|\varepsilon|_e &\leq 0.
\end{aligned}
\end{equation}
Writing (\ref{42-A}) as a quadratic form in terms of the expended vector $[|\varepsilon|_e^{\top},|\hat{\eta}|_e^{\top}]^{\top}$ yields  
\begin{equation}\label{Matrix-C}
\begin{aligned}
	c\overbrace{\left[\begin{array}{cc} 
		(1-c_1)	\delta_{{A}_0}\bar{\mathbf{A}}_0 &0\\
		0&0
		\end{array}\right] }^{\mathbf{Q}_{c_1}} &\\
	+\underbrace{\left[\begin{array}{cc} 
		0& |\mathbf{P}_\varepsilon|_e^{\top}\\
		|\mathbf{P}_\varepsilon|_e & (1-c_1)\delta_{{A}_c}\bar{\mathbf{A}}_c
		\end{array}\right]}_{\mathbf{Q}_{c_2}} & \leq 0 . 
	\end{aligned}
\end{equation}	
Note that the block matrix in the upper left corner of $\mathbf{Q}_{c_1}$ and the block matrix in the lower right corner of $\mathbf{Q}_{c_2}$ are SND, and both $\mathbf{Q}_{c_1}$ and $\mathbf{Q}_{c_2}$ are symmetric, so it is possible to select
\begin{equation}\label{c}
c=\max\left\{\sigma\left(\mathbf{Q}_{c_2},-\mathbf{Q}_{c_1}\right)\right\} ,
\end{equation}
such that (\ref{Matrix-C}) holds, \textcolor{black}{which further guarantees that (\ref{42}) holds.}

\textbf{\textit{Step 2:}}    $b$ is selected to ensure that the following inequality is satisfied with $\kappa_{a}$ being an arbitrary  positive constant:
\begin{equation}\label{47}
\hspace{-1cm}\begin{aligned}
b\textcolor{black}{c_1}\left(\hat\eta^{\top}\hspace{-0.1cm}\left(\mathbf{P}_c\mathbf{A}_c+\mathbf{A}_c^{\top}\hspace{-0.05cm}\mathbf{P}_c\right)\hat\eta +c\varepsilon^{\top}\hspace{-0.05cm}\left(\mathbf{P}_0\mathbf{A}_0+\mathbf{A}_0^{\top}\mathbf{P}_0\right)\varepsilon\right)&\\+\left(\hat\eta^{\top}\hspace{-0.1cm}(\mathbf{P}_c\mathbf{D}+\mathbf{D}\mathbf{P}_c)\hat\eta +c\varepsilon^{\top}\hspace{-0.1cm}(\mathbf{P}_0\mathbf{D}+\mathbf{D}\mathbf{P}_0)\varepsilon\right) &\\
\textcolor{black}{+2b \hat{\eta}  ^{\top} \hat\eta} +\kappa_{a}\left(\hat\eta^{\top}\mathbf{P}_c\hat\eta+c\varepsilon^{\top}\mathbf{P}_0\varepsilon\right)&\leq 0. 
\end{aligned}\hspace{-1cm}
\end{equation}
\textcolor{black}{Writing (\ref{47}) as a quadratic form in terms of   $[ \varepsilon ^{\top}, \hat\eta ^{\top}]^{\top}$ yields,}
\begin{equation}\label{Matrix-b}
 \hspace{-1.5em}\begin{aligned}
	b\overbrace{\left[\begin{array}{cc} 
		\hspace{-0.1cm}c_1c(\mathbf{P}_0\mathbf{A}_0+\mathbf{A}_0^{\top}\hspace{-0.05cm}\mathbf{P}_0 )\hspace{-0.3cm} &\hspace{-0.05cm}0\hspace{-0.05cm}\\
		\hspace{-0.1cm}0  \hspace{-0.05cm}&\hspace{-0.3cm}c_1(\mathbf{P}_c\mathbf{A}_c+\mathbf{A}_c^{\top}\hspace{-0.05cm}\mathbf{P}_c )+2\mathbf{I}_{nN}\hspace{-0.05cm}
		\end{array}\right] }^{\mathbf{Q}_{b_1}}&\\
	\hspace{-0.05cm}+\hspace{-0.05cm}\underbrace{\left[\begin{array}{cc} 
		\hspace{-0.1cm}c(\mathbf{P}_0\mathbf{D}+\mathbf{D}\mathbf{P}_0) +c\kappa_a \mathbf{P}_0\hspace{-0.3cm}& \hspace{-0.07cm}0  \hspace{-0.07cm}\\
		\hspace{-0.1cm} 0 \hspace{-0.07cm} &\hspace{-0.3cm} (\mathbf{P}_c\mathbf{D}+\mathbf{D}\mathbf{P}_c) +\kappa_a \mathbf{P}_c \hspace{-0.07cm}
		\end{array}\right]}_{\mathbf{Q}_{b_2}}  &  \hspace{-0.05cm}\leq\hspace{-0.05cm} 0 .
	\end{aligned} \hspace{-2em}
\end{equation} 
Note that $\mathbf{Q}_{b_1}$ and $\mathbf{Q}_{b_2}$ are both symmetric matrices and $-\mathbf{Q}_{b_1}$ is SPD. According to Lemma \ref{lemma-matrix} and noting that $b\geq T$, it is sufficient to choose
\begin{equation}\label{b}
b=\max\left\{\sigma\left(\mathbf{Q}_{b_2},-\mathbf{Q}_{b_1}\right),T\right\} 
\end{equation} 
such that (\ref{Matrix-b}) holds, which further guarantees that (\ref{47}) holds.

\textcolor{black}{\textbf{\textit{Step 3:}} Here we aim to verify that the following inequality  holds for some positive constant $\kappa_b$,}
\begin{equation}\label{53} 
\textcolor{black}{\begin{aligned}
	2bc\varepsilon^{\top}\mathbf{P}_0\Phi  \leq \frac{\kappa_b}{\alpha(\tau)} V.
	\end{aligned}}
\end{equation} 
By recalling (\ref{29}) and using Assumption \ref{ass-1} and the fact $r\geq 1$, we have  
\begin{equation}\label{Phi}
|\Phi|_e\leq_e \frac{1}{r}\mathbf{A}_\varrho(|\hat\eta|_e+|\varepsilon|_e),
\end{equation}
where $\mathbf{A}_{\varrho}= \operatorname{diag}\{\mathbf{A}_{\varrho_1},\cdots,\mathbf{A}_{\varrho_N}\}$ with $\mathbf{A}_{\varrho_k} (k=1,\cdots,N)$ being defined in (\ref{Arhok}). Accordingly, \textcolor{black}{the effects of uncertainties on the Lyapunov inequality (\ref{dV-new}) are bounded by}
\begin{equation}\label{52}
\begin{aligned}
2bc\varepsilon^{\top}\mathbf{P}_0\Phi \leq \frac{2c }{\alpha(\tau)}|\varepsilon|_e^{\top}|\mathbf{P}_0\mathbf{A}_\varrho|_e(|\hat\eta|_e +|\varepsilon|_e) .
\end{aligned}
\end{equation}  
\textcolor{black}{Let $\bar{\mathbf{P}}_c=\operatorname{diag}\{\mathbf{P}_c\}$ and $\bar{\mathbf{P}}_0=\operatorname{diag}\{\mathbf{P}_0\}$ be two positive definite diagonal matrices. Hence, from Lemma \ref{lemma-matrix},  there exist two positive constants $ \kappa_b$ and $\tilde\kappa_b$, such that,}
\begin{equation} 
\begin{aligned}
	\tilde\kappa_b\overbrace{\left[\begin{array}{cc} 
		\hspace{-0.05cm}-\bar{\mathbf{P}}_0 &0\\
		\hspace{-0.05cm}0&-\bar{\mathbf{P}}_c
		\end{array}\right]}^{\mathbf{Q}_{\kappa_a}} &\\
	+\underbrace{c\left[\hspace{-0.15cm}\begin{array}{cc}    |\mathbf{P}_0|_e\mathbf{A}_\varrho+\mathbf{A}_\varrho^{\top}|\mathbf{P}_0|_e&   |\mathbf{P}_0|_e\mathbf{A}_\varrho\\
		\mathbf{A}_\varrho^{\top}|\mathbf{P}_0|_e& 0 
		\end{array}\hspace{-0.15cm}\right]} _{\mathbf{Q}_{\kappa_b}}& \leq 0 , 
	\end{aligned}
\end{equation}
\begin{equation}\label{Matrix-K}
\textcolor{black}{\begin{aligned}
	\kappa_b\underbrace{\left[\begin{array}{cc} 
		-c {\mathbf{P}}_0 &0\\
		0&- {\mathbf{P}}_c
		\end{array}\right]}_{\mathbf{Q}_{\kappa_c}} 
	+\underbrace{\left[\begin{array}{cc} 
		\tilde\kappa_b\bar{\mathbf{P}}_0& 0\\
		0& \tilde\kappa_b\bar{\mathbf{P}}_c
		\end{array}\right]}_{\mathbf{Q}_{\kappa_d}} \leq 0 .
	\end{aligned}}
\end{equation}
\textcolor{black}{Specifically, the suitable $ \kappa_b$ and $\tilde\kappa_b$ are}
\begin{equation}\label{kappa-1}
\textcolor{black}{\begin{aligned} 
	\kappa_b &=\max\{\sigma(\mathbf{Q}_{\kappa_d},-\mathbf{Q}_{\kappa_c})\},~ \\
	\tilde\kappa_b&=\max\{\sigma(\mathbf{Q}_{\kappa_b},-\mathbf{Q}_{\kappa_a})\}.
	\end{aligned}\hspace{-1em}}
\end{equation}  
\textcolor{black}{Similar to (\ref{kappa-2}),  by means of (\ref{Matrix-K}) and the fact $ \eta^{\top}\bar{\mathbf{P}}_c \eta=| \eta|_e^{\top}\bar{\mathbf{P}}_c| \eta|_e$ and $\varepsilon^{\top}\bar{\mathbf{P}}_0\varepsilon=|\varepsilon|_e^{\top}\bar{\mathbf{P}}_0|\varepsilon|_e$, one can conclude that,}
\begin{equation}\label{58-1}
\begin{aligned}
	&\frac{2c }{\alpha(\tau)}|\varepsilon|_e^{\top}|\mathbf{P}_0\mathbf{A}_\varrho|_e(|\hat\eta|_e  \\ \leq & \frac{{\kappa}_b}{\alpha(\tau)} \left( \hat\eta^{\top}\mathbf{P}_c\hat\eta+c\varepsilon^{\top}\mathbf{P}_0\varepsilon \right) .
	\end{aligned} 
\end{equation} 
\textcolor{black}{Recalling (\ref{35}) and (\ref{52}), we know that (\ref{53}) holds. Combining (\ref{42}), (\ref{47}), and (\ref{53}), it can be verified that (\ref{dV-final}) holds.  
	Similar to (\ref{vv}) and (\ref{54-1}), we have from Lemma \ref{lemma-2} that}  
\begin{equation}\label{54-2}
\textcolor{black}{V (t)\leq \left(\frac{{T}-t}{{T}}\right)^{\tilde\kappa}\tilde MV (0) ,~t\in[0,{T}),}
\end{equation}  
where $\tilde\kappa=\kappa_a-\kappa_b/\alpha(\tau_1)$ and $\tilde M=e^{T\kappa_b\tau_1}$ is a finite number with $0<\tau_1<\infty$. Additionally, by recalling (\ref{35}), we get
\begin{equation}\label{54-1-2}
|\eta(t)|+|\varepsilon(t)|\leq \frac{2\sqrt{V(t)}}{\sqrt{\min\{\lambda_{\min}(\mathbf{P}_c),c\lambda_{\min}(\mathbf{P}_0)\}}} .
\end{equation} 
\textcolor{black}{Similar to the analysis in footnote 3, the above estimate does not lead to conservative controller parameter selections.}   With $\mathbf{\Lambda}_r = \operatorname{diag}\{r^{n},r^{n-1},\cdots,r\}$ and the fact $r \geq  1$, it holds that 
\begin{equation}\label{55-1-2}
\textcolor{black}{\hspace{-0.2em}\begin{aligned}
 |x_k-x_0|&\leq \frac{1}{r} (|\eta|+|\varepsilon|)\\&\leq  \frac{2\Big(\frac{(T-t)^{\tilde{\kappa}+2}}{T^{\tilde{\kappa}}}\Big)^{\frac{1}{2}} \sqrt{\tilde{M}V(0)}}{b\sqrt{\min\{\lambda_{\min}(\mathbf{P}_c),c\lambda_{\min}(\mathbf{P}_0)\}}},  
	\end{aligned}\hspace{-2em}}
\end{equation}
\textcolor{black}{establishing the same for $|\hat{x}_k-\hat{x}_0|$.
	Hence, it follows that $x_k-x_0 \in\mathcal{L}_\infty[0,T)$, $\hat{x}_k-\hat{x}_0 \in\mathcal{L}_\infty[0,T)$, $\lim_{t\rightarrow T}|x_k-x_0|=0$, and $\lim_{t\rightarrow T}|\hat{x}_k-\hat{x}_0|=0$.  By recalling (\ref{32}), we know that the designed control inputs satisfy $u_k \in\mathcal{L}_\infty[0,T)$.    
	The proof is completed. $\hfill\blacksquare$}

\begin{remark}
	\textcolor{black}{Due to each agent being equipped with sensors of different sensitivities $\theta_k(t)$, addressing these unknown parameters individually is not straightforward.  The proposed solution is to aggregate all unknown sensitivities into a matrix for unified processing, which is technically feasible but may lead to more conservative design parameters (i.e., smaller allowable sensitivity errors $\Delta_{\theta_k}$). This is because we only incorporate the parameters of the poorest-performing sensor into the design of the entire distributed controller. As the system scale increases (i.e.,  as $N$ becomes larger), the value of $1/|\mathbf{P}_c\mathbf{A}_g|$ typically decreases accordingly. According to (\ref{42-2}), it can be inferred that in such a scenario, sensors with higher precision are needed for the proposed algorithm to be applicable.}
\end{remark}

\subsection{Practical prescribed-time consensus tracking}

%正式陈述在定理3中

Note that $\alpha(t)$ in Theorem \ref{theorem-3}   escapes to $\infty$ as $t$ approaches the terminal time $T$. \textcolor{black}{Although the existing literature (see, for instance, \cite{2020-SCL-Chenxiandong}) has preliminarily explored applying the prescribed-time controller to the entire time domain by imposing saturation on the time-varying gain before $t$ approaches the terminal time, its viability has not been thoroughly analyzed. In this subsection, we provide a quantitative relationship between the maximum saturation level of the time-varying gain $\alpha(t)$ and the terminal time $T$.}

\textcolor{black}{In order to facilitate the association of the solution under bounded time-varying feedback with the control scheme introduced earlier, we redefine $T=t_f+\delta$, with $t_f$ and $b$ being two arbitrary positive constants. Then we have the following practical prescribed-time protocol whose proof is provided for completeness.}

\begin{theorem}\label{corollary-2}
	Suppose that the graph \textcolor{black}{${\mathcal{G}}$} is connected and Assumption \ref{ass-1} is satisfied with {known} growth rates $\varrho_{k,j}$. \textcolor{black}{If maximum value of the allowable sensitivity error in Assumption \ref{ass-2} satisfies} 
	\begin{equation}\label{allowable-A}
	\textcolor{black}{0<\max \{\Delta_{\theta_1},\cdots,\Delta_{\theta_N}\}\leq  \frac{1}{ |\mathbf{P}_c\mathbf{A}_g|},}
	\end{equation}	
	and the time-varying gain of the controller and the observer is replaced by $r=b\alpha_0$, with
	\begin{equation}\label{alpha0-t}
	\textcolor{black}{\alpha_0(t)=\left\{ \begin{array}{ll} \vspace{0.1cm}
		\frac{1}{t_f+\delta-t}  ,~t\in\left[0,t_f \right), \\
		\frac{1}{\delta},~~~ ~~~~~ t\in\left[t_f,\infty \right),
		\end{array}\right.}
	\end{equation} 
	\textcolor{black}{where $1/\delta$, denotes the maximum saturation level of the time-varying gain, can be selected as any positive constant,} then we can find a proportional coefficient $b$ such that: \textcolor{black}{$1)$ the  closed-loop dynamic system  composed of (\ref{observer}), (\ref{30}),  and (\ref{32}) is asymptotically stable on $t\in[0,\infty)$; and $2)$ the tracking error $e_k=x_k-x_0$ converge to a compact set $\Omega$ as $t\rightarrow t_f$, where}
	\begin{equation}\label{omega}  \textcolor{black}{\Omega\hspace{-0.1em}=\hspace{-0.1em}\left\{\hspace{-0.1em}|e_k|\in\mathbb{R}\Big| |e_k|\leq\frac{2\delta  \sqrt{V(0)}({\delta}/{t_f+\delta})^{{\gamma}/{2}}}{b\sqrt{\min\{\lambda_{\min}(\mathbf{P}_c),c\lambda_{\min}(\mathbf{P}_0)\}}} \right\}\hspace{-0.2em},}
	\end{equation} 
	\textcolor{black}{with  $\gamma>0$ being an arbitrarily large  parameter, and $V(0)$ being the initial value of $V$ defined in (\ref{35}).}
\end{theorem}	

\textit{Proof:} The proof is implemented by two stages. Firstly, for $t\in[0,t_f)$, functions $\alpha_0(t)$ and $\alpha(t)$ are equivalent. According to the proof of Theorem \ref{theorem-3}, let $V(t)=\hat\eta^{\top}\mathbf{P}_c\hat\eta+c\varepsilon^{\top}\mathbf{P}_0\varepsilon,$ \textcolor{black}{we know that choosing  $c$ and $b$ according to (\ref{c}) and (\ref{b}) with $\kappa_a$  being}
\begin{equation} \label{Kp}
\textcolor{black}{\kappa_a>\frac{\kappa_1}{\inf\{\alpha_0\}}= \kappa_b(t_f+\delta),}
\end{equation}
\textcolor{black}{where  $\kappa_b$ is  calculated by (\ref{allowable-A}), can make the derivative of $V$ satisfies} 
\begin{equation}\label{75}
\textcolor{black}{\frac{dV}{d\tau}\leq -\left(\kappa_a-\frac{\kappa_b}{\alpha(\tau)}\right)V\leq -\gamma V,}
\end{equation}
\textcolor{black}{where $\gamma=\kappa_a-\kappa_b(t_f+\delta)>0$.
	This indicates that  $V(t)\in\mathcal{L}_{\infty}[0,t_f)$ and $V(\tau)\leq e^{-\gamma \tau}V(0)$, which further implies that, by recalling $\tau=\ln(T/(T-t))$ and $T=t_f+\delta$,}
\begin{equation} \label{71}
\textcolor{black}{\lim_{t\rightarrow t_f}V(t)\leq \left(\frac{\delta}{t_f+\delta}\right)^{\gamma}V(0).} 
\end{equation} \textcolor{black}{According to (\ref{55-1-2}), we have}
\begin{equation}\label{59-1}
\lim_{t\rightarrow t_f}|e_k|\leq \frac{2\delta  \sqrt{V(0)}\left(\frac{\delta}{t_f+\delta}\right)^{\frac{\gamma}{2}}}{b\sqrt{\min\{\lambda_{\min}(\mathbf{P}_c),c\lambda_{\min}(\mathbf{P}_0)\}}} .
\end{equation} 

Secondly, for $t\in[t_f,\infty)$, we have $\alpha_0\equiv1/\delta$, then $r=b/\delta$ becomes a constant rather than a time-varying parameter.  Therefore, the dynamics of $\eta$ and $\varepsilon$ in (\ref{33}) can be simplified to
\begin{equation}\label{64}
\begin{aligned}
\dot{\hat\eta} &=r\left(\mathbf{A}_c\hat\eta+G^{\top}\otimes \left(\textcolor{black}{\mathbf{I}_\theta}\Psi_{\varepsilon}+\textcolor{black}{(\mathbf{I}_\theta-\mathbf{I}_N)}\Psi_{\hat\eta}\right)\right) , \\
\dot\varepsilon &=r\left(\mathbf{A}_0\varepsilon+\Phi\right)   ,
\end{aligned}
\end{equation} 
\textcolor{black}{Similar to the treatment given in (\ref{23})-(\ref{49-2}),} the derivative of $V$ w.r.t. $t$ along (\ref{64}) can be calculated as 
\begin{equation}\label{dV-O}
\begin{aligned}
\dot{V} 
\leq&~ \frac{b}{\delta}   \hat\eta^{\top}\hspace{-0.1cm}\left(\mathbf{P}_c\mathbf{A}_c\hspace{-0.05cm}+\hspace{-0.05cm}\mathbf{A}_c^{\top}\mathbf{P}_c\right)\hspace{-0.05cm} \hat\eta \hspace{-0.05cm}+\hspace{-0.05cm}\frac{bc}{\delta}  \varepsilon^{\top}\hspace{-0.1cm}\left(\mathbf{P}_0\mathbf{A}_0\hspace{-0.05cm}+\hspace{-0.05cm}\mathbf{A}_0^{\top}\mathbf{P}_0\right)\hspace{-0.05cm}\varepsilon\\
&~+  \frac{2bc}{\delta}\varepsilon^{\top}\mathbf{P}_0\Phi + \frac{2b}{\delta}\left( \hat{\eta}  ^{\top}  \hat{\eta} +  |\hat{\eta}|_e^{\top}\textcolor{black}{|\mathbf{P}_\varepsilon|_e}|\varepsilon|_e\right).
\end{aligned}
\end{equation}
Analogous to the analysis in the proof of Theorem \ref{theorem-3}, the design purpose is to find parameters $b$ and $c$ such that (\ref{dV-O}) can be reduced to \textcolor{black}{$\dot{V}\leq -\gamma^* V$ with $\gamma^*>0$}. \textcolor{black}{From (\ref{42})-(\ref{c}), we know that select  $
	c=\max \{\sigma (\mathbf{Q}_{c_2},-\mathbf{Q}_{c_1} ) \}$
	can ensure that, for $c_1\in(0,1)$,} 
\begin{equation}\label{dV-O-1}
\textcolor{black}{\begin{aligned}
	(1-c_1)\frac{b}{\delta}  \left( \hat\eta^{\top} (\mathbf{P}_c\mathbf{A}_c\hspace{-0.05cm}+\hspace{-0.05cm}\mathbf{A}_c^{\top}\mathbf{P}_c)  \hat\eta \right. &\\  \left.+c \varepsilon^{\top} (\mathbf{P}_0\mathbf{A}_0\hspace{-0.05cm}+\hspace{-0.05cm}\mathbf{A}_0^{\top}\mathbf{P}_0) \varepsilon\right)
	+   \frac{2b}{\delta}  |\hat{\eta}|_e^{\top}\textcolor{black}{|\mathbf{P}_\varepsilon|_e}|\varepsilon|_e &\leq 0.
	\end{aligned}}
\end{equation} 
\textcolor{black}{Next, in accordance with (\ref{53})-(\ref{58-1}), we know that choose $
	\tilde\kappa_b=\max\{\sigma(\mathbf{Q}_{\kappa_b},-\mathbf{Q}_{\kappa_a})\},$ and $\kappa_b=\max\{\sigma(\mathbf{Q}_{\kappa_d},-\mathbf{Q}_{\kappa_c})\}$ can ensure that}
\begin{equation}\label{ccc} 
\textcolor{black}{2bc\varepsilon^{\top}\mathbf{P}_0\Phi  \leq \frac{\kappa_b}{\alpha(\tau)} V.} 
\end{equation}
\textcolor{black}{Since $r(t)$ is always a constant (i.e., $\dot{r}\equiv 0$) after $t\geq t_f$, the two SPD matrices  produced by $\dot{r}$ appearing in $\mathbf{Q}_{b_2}$ (i.e. $\mathbf{P}_0\mathbf{D}+\mathbf{D}\mathbf{P}_c$ and $\mathbf{P}_0\mathbf{D}+\mathbf{D}\mathbf{P}_0$ in (\ref{Matrix-b})) can be removed when recalculating $b$ in this part. Namely, picking} 
\begin{equation} \label{B}
\textcolor{black}{b=\max \left\{\sigma (\mathbf{Q}_{b_2^{*}},-\mathbf{Q}_{b_1} ),t_f+\delta \right\}} 
\end{equation} 
\textcolor{black}{with $\mathbf{Q}_{b_2^{*}}=\operatorname{diag}\{c\kappa_a\mathbf{P}_0,\kappa_a\mathbf{P}_c\}$ can yield}
\begin{equation}\label{47-1}
\textcolor{black}{\begin{aligned}
	\frac{bc_1}{\delta}\left(\hat\eta^{\top}  (\mathbf{P}_c\mathbf{A}_c+\mathbf{A}_c^{\top} \mathbf{P}_c ) \hat\eta \right. & \\ \left.+c\varepsilon^{\top}  (\mathbf{P}_0\mathbf{A}_0+\mathbf{A}_0^{\top}\mathbf{P}_0 )\varepsilon\right)
	+\frac{2b}{\delta} \hat{\eta}  ^{\top} \hat\eta +\kappa_{a}V& \leq 0.
	\end{aligned}}
\end{equation}  
\textcolor{black}{Combining (\ref{dV-O-1}), (\ref{ccc}),  and (\ref{47-1}), we know that (\ref{dV-O}) satisfies}
\begin{equation}\label{dV-O-2} 
\textcolor{black}{\dot{V} 
	\leq -\frac{\kappa_{a}}{\delta} V+\kappa_bV.} 
\end{equation}
\textcolor{black}{From (\ref{Kp}), we know $\kappa_a=\kappa_b(t_f+\delta)>\kappa_b\delta$, which indicates that} 
\begin{equation}\label{VVV}
\textcolor{black}{\dot{V}\leq -\gamma^*V,~t\in[t_f,\infty),}
\end{equation} 
\textcolor{black}{where $\gamma^*=\kappa_a/\delta-\kappa_b>0$. Solving (\ref{VVV}) and recalling (\ref{71}) gives
	\begin{equation}
	V(t)\leq e^{-\gamma^* t}V(t_f)\leq e^{-\gamma^* t}\left(\frac{\delta}{t_f+\delta}\right)^{\gamma}V(0).
	\end{equation}	
	Therefore, the conclusion of $V\in\mathcal{L}_\infty[t_f,\infty)$ and $\lim_{t\rightarrow \infty}V=0$ can be obtained under the selection of $c$ and $b$ in (\ref{c}) and (\ref{b}).} It follows from (\ref{54}) that $\hat\eta\in\mathcal{L}_\infty[0,\infty)$, $\varepsilon\in\mathcal{L}_\infty[0,\infty)$, and $\lim_{t\rightarrow \infty}|\hat\eta|=\lim_{t\rightarrow \infty}|\varepsilon|=0$.  
Recalling (\ref{55-1}), one can conclude that $e_k  \in\mathcal{L}_\infty[0,\infty)$ and   $\lim_{t\rightarrow \infty}|e_k |=0$. By recalling (\ref{32}), it also holds that  $u_k \in\mathcal{L}_\infty[0,\infty)$.  
This completes the proof.   
$\hfill\blacksquare$

\begin{remark}
	\textcolor{black}{Since the parameter $\delta$ in (\ref{59-1}) can be chosen to be sufficiently small, such as $\delta=0.001$, and the power of $\delta$ (i.e., $\gamma$) can be chosen to be sufficiently large by increasing $\kappa_a$, we can consider $ \Omega$ defined in (\ref{omega}) as a compact set with an adjustable radius. Notice that all parameters including $\delta,\gamma,t_f,b,c,\lambda_{\min}(\mathbf{P}_c)$ and $\lambda_{\min}(\mathbf{P}_0)$ are known, so the radius of the compact set $\Omega$ can be derived from the initial condition $V(0)$. In this subsection, the selection of $\kappa_a$ needs to satisfy the quantitative relationship given by (\ref{Kp}) rather than any positive number  allowed by (\ref{47}).}
\end{remark}  

\begin{remark}%方法比li和chen更简单因为只用了一个增益;
	%不同于文献A和文献B使用两个增益来分别处理系统非线性和未知测量灵敏度，定理和推论2只使用了一个时变增益来同时处理闭环系统中所有的不确定性，这样做的好处在于在不增加控制器复杂度的前提下加快了收敛速度。同时，整个参数设计过程是在矩阵束框架下进行的，正如在例子1中演示的那样，这样做的好处是保证设计参数的低保守性。
	\textcolor{black}{Different from the approach in \cite{2018-chen-tac} and \cite{2020-auto-liwuquan} which use dual-domination gains to deal with system nonlinearities and unknown sensor sensitivities, this paper uses only a time-varying gain to comprehensively handle all uncertainties in the closed-loop system. The advantage of doing so is that it accelerates the convergence rate without increasing the controller complexity. In particular, the entire parameter design process is conducted within the matrix pencil based framework, as demonstrated in   \cite{2020-NYU-IFAC,2020-CDC-matrix-pencil,2022-scl-matrix-pencil}, which ensures low conservatism of the design parameters. In addition,  the proposed method ensures that all states converge to a compact set within the prescribed time. This extends the practical prescribed-time control scheme presented in \cite{2022-caoye} and \cite{2023-tanyan}, which only guarantees the convergence of the outputs to a compact set within the prescribed time.}
\end{remark}

%\begin{remark}\textcolor{red}{Compared with the typical works on finite- and fixed-time consensus where the estimation approaches were used with conservation (see, for instance, \cite{2018-IJC-Ning-finite=fix,2018-TII-Zuo-fix-finite,2019-TAC-Ning-prescribed}), in this paper the settling time can be off-line pre-assigned according to task requirements irrespective of initial conditions and any other design parameter. Besides, we investigate the consensus problem of MASs under output-feedback, which is fundamentally different from the state-of-the-art works on consensus in MASs under state-feedback or accurate output-feedback (see, for instance, \cite{2018-Auto-Wang,2018-Cyber-Wang,2020-SCL-Chenxiandong}), as only inaccurate and non-differentiable output signals are available for control design.} \end{remark}

\section{Simulations}\label{sec-simulation}
To verify the effectiveness of our prescribed-time  protocols, we consider a network consisting of a leader and four followers. \textcolor{black}{Each agent  is modeled as a single-link robot manipulator, described by the Lagrangian equation} 
\begin{figure*}[!h]	 
	\centering  
	\includegraphics[width=7 in]{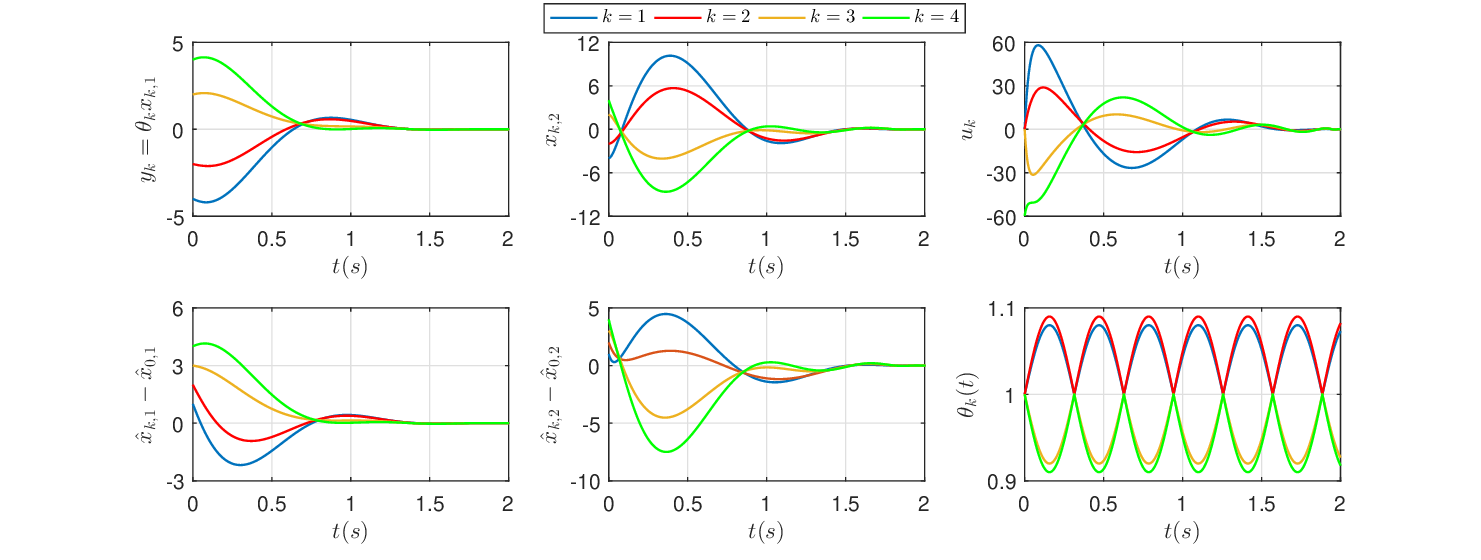} 
	\caption{\textcolor{black}{States, control, and observer errors for MASs with the settling time $t_f+\delta=2$s  under output feedback.}}\label{fig-state-4}
\end{figure*}  
\begin{equation}\label{model-robot}
J_k\ddot{q}_k+B_k\dot{q}_k+m_kgh_k\sin(q_k)=u_k,~k=0,1\cdots,4
\end{equation}
where $q_k$ and $\dot{q}_k$ are the general position and velocity, respectively, $J_k$ denotes the rotation inertia, $B_k$ is the damping coefficient, $h_k$ is the length from the axis of joint to the mass center, $m_k$ is the mass of the link, and $g$ is the acceleration of gravity. Furthermore, $J_0=J_4=8.5$, $J_1=J_2=J_3 =10$, $B_0=B_3=1.4$, $B_1=B_2=B_4=1.6$, $h_0=h_1=1$, $h_2=0.8$, $h_3=h_4=1.2$, $m_0= m_2=1.3$, $m_1=m_3=m_4=1$, $g=9.8$.   \textcolor{black}{Laplacian matrix} \textcolor{black}{of graph $\bar{\mathcal{G}}$ is given as follows}
\begin{equation*}
\textcolor{black}{\begin{aligned}
	\mathcal{L}_{\bar{\mathcal{G}}} =\left[\begin{array}{cccc} 
	2 & -1&0& 0\\
	-1 & 2&-1&0\\
	0 & -1&2& -1\\
	0 & 0&-1&1
	\end{array}\right].  
	\end{aligned}}
\end{equation*}

Our control task is to design a distributed controller such that all the followers' states can track the states of the leader within a pre-given time. Define $x_k=[x_{k,1},x_{k,2}]^{\top}=[J_kq_k,J_k\dot{q}_k]^{\top}$ and $y_k=J_kq_k$, then (\ref{model-robot}) can be rewritten as the same form as $\dot{x}_k=\mathbf{A}x_k+Bu_k+F_k$, $y_k=\textcolor{black}{\theta_k(t)}Cx_k$ with $\mathbf{A}=[0 ,1; 0 ,0]$, $B=[0;1]$, $C=[1,0]$, and $F_k=[0;- {B_k}x_{k,2}/{J_k}-m_kgh_k\sin({x_{k,1}}/{J_k})]$. 
It can be verified that Assumption \ref{ass-1} is satisfied with $\varrho_{k,1}=1.8$ and $\varrho_{k,2}=0.19$ for $k=1,2,3,4$. In the simulation, the initial conditions are set to $x_0(0)=(0;0)$, $x_1(0)=(1;1)$, $x_2(0)=(2,2)$, $x_3(0)=(3,3)$, $x_4(0)=(4;4)$,  the input of the leader agent is set to \textcolor{black}{$u_0(t)=0$}, and  the sensor sensitivity $\theta_k(t)$ is set to \textcolor{black}{$\theta_k(t)=1+\varkappa_k|\sin(10t)|$ ($k=0,1,2,3,4$) with $\varkappa_k\in (-0.1,0.1)$}.

For the purpose of verifying the effectiveness of the protocol stated in Theorem \ref{theorem-3}, we use the distributed prescribed-time output-feedback protocols given in Theorem \ref{theorem-3} for simulation, i.e., \textcolor{black}{$
u_k=-K\mathbf{\Lambda}_r (\sum_{j=1}^{4}a_{k,j}(\hat{x}_k-\hat{x}_j)+b_k(\hat{x}_k-\hat{x}_0) )$} where $\mathbf{\Lambda}_r=\operatorname{diag}\{r^{2}, r\}$, $r=b\alpha(t) $ with $\alpha$ being defined in (\ref{alpha-t}).
Besides, the distributed observer is given by
\begin{equation*} 
\textcolor{black}{\begin{aligned}
	\dot{\hat{x}}_{k}=\mathbf{A}\hat{x}_{k}+Bu_k+r^{n+1}\mathbf{\Lambda}_r^{-1}G(y_k-\hat{x}_{k,1}),
	\end{aligned}}
\end{equation*}
where  $K=[8,9]$ and  \textcolor{black}{$G=[2,2]$. Let \textcolor{black}{$a_0=0,a_1=0.08,a_2=0.09,a_3=-0.08,a_4=-0.09$, and $c_1=0.9$}, we obtain two \textcolor{black}{$8\times 8$-dimensional} SPD matrices $\mathbf{P}_c$ and $\mathbf{P}_0$ in accordance on the basis of (\ref{58}). Let $\kappa_a=0.001$, $\delta=0.02$  and $t_f=1.98$, according to (\ref{59-A}), (\ref{c}), (\ref{b}),  and   the expressions for $\mathbf{P}_c$, $\mathbf{P}_0$, $\bar{\mathbf{P}}_c$, $\bar{\mathbf{P}}_0$, $\mathbf{D}$, ${\mathbf{A}}_g$, ${\mathbf{A}}_\varrho$, ${\mathbf{A}}_c$, ${\mathbf{A}}_0$, $\bar{\mathbf{A}}_c$, and $\bar{\mathbf{A}}_0$, the design parameters can be calculated as $\delta_{{A}_c}=\delta_{{A}_0}=1$,   
	$c = \max \{\sigma (\mathbf{Q}_{c_2},-\mathbf{Q}_{c_1} ) \}=565.79, $ and $
	b =\max \{\sigma (\mathbf{Q}_{b_2^{*}},-\mathbf{Q}_{b_1} ),t_f+\delta \} =4.45.$}
\textcolor{black}{In addition, it can be checked that the sensitivity error under consideration is allowable since  $\max \{\Delta_{\theta_1},\cdots,\Delta_{\theta_4}\}\leq 0.1\leq{1}/{|\mathbf{P}_c\mathbf{A}_g|}= 0.1015$}.   

The simulation results are presented in Fig. \ref{fig-state-4}, which shows that the followers' states (with different initial values) tracks the leader's states within the pre-given time, and the control input signals are uniformly bounded over the prescribed time interval.  In addition, it can be seen that all observer errors converge to zero within the prescribed time, and the sensor sensitivity $\theta_k$ for each agent is different. These findings confirm our theoretical predictions and affirm the efficacy of the proposed prescribed-time control strategy.

\section{Conclusions}\label{sec-conclusion}
In this paper, we have tackled the leader-following consensus problem by designing distributed state-feedback and output-feedback controllers using local information (might be inaccurate) to ensure that all the agents track the leader's states within a prescribed time. 
Furthermore, \textcolor{black}{we robustify the standard matrix pencil-based design procedure in the presence of multiplicative measurement noise and develop its distributed formulation to select the proportional coefficient of the time-varying control gain,} and adopt the time-axis transformation to convert the original system operating over a finite time interval into one that operates over an infinite time interval. 
In this way, the prescribed-time control problem can be recast into the asymptotic control problem, which simplifies the control design and stability analysis.
Additionally, we achieve practical prescribed-time consensus for MAS systems by embedding bounded time-varying gain without sacrificing the ultimate control accuracy. 
In particular, the prescribed-time and the practical prescribed-time leader-following controllers (including the observer) have the same structure. 
The latter can be obtained directly by applying saturation to the unbounded gain of the former, and we quantitatively discuss how to choose this saturation constant.  
The simulation results verify the effectiveness of the proposed method. 
Based on these results, it is of interest to further study prescribed-time formation control for MASs with more general topologies and topology attacks.

\appendix

\subsection{Proof of Proposition 1}  
Taking the derivative of (\ref{eta_k})   along the trajectory of (\ref{model}) yields 
\begin{equation} \label{56}
\begin{aligned}
\dot{{\eta}}_k =r\left(\mathbf{A}{\eta}_k+\textcolor{black}{B u_k} +\Phi_k\right)+\frac{\dot{r}}{r}\mathbf{D}_c{\eta}_k .
\end{aligned}
\end{equation} 
Applying the time transformation on the foregoing ODE, we can rewrite (\ref{56}) as its equivalent form on the time domain $[0,\infty)$,
\begin{equation} \label{app-2}
\begin{aligned}
 \frac{d{{\eta}}_k}{d\tau} =\frac{dt}{d\tau}r\left(\mathbf{A}{\eta}_k+\textcolor{black}{B u_k} +\Phi_k\right)+\frac{dr}{d\tau}\frac{1}{r}\mathbf{D}_c{\eta}_k,
\end{aligned}
\end{equation} 
Recalling that $r=b\alpha(\tau)=b ({d\tau}/{dt})=be^{\tau}/t_f$, we have $$\frac{dt}{d\tau}r=b,~~\frac{dr}{d\tau}\frac{1}{r}=1.$$ Accordingly, (\ref{app-2}) can be continued as
\begin{equation*} 
\begin{aligned}
 \frac{d{ {\eta}_k}}{d\tau}=b\left(\mathbf{A} {\eta}_k+\textcolor{black}{B u_k}+\Phi_k\right)+\mathbf{D}_c {\eta}_k .
\end{aligned}
\end{equation*} 
The proof is completed. $\hfill\blacksquare$

%\begin{lemma}[{[\citen{2015-Auto-Zhang}]}]\label{lemma-1}
%	Two row vectors $K=[k_1,\cdots,k_n]$ and $G=[g_1,\cdots,g_n]$ with $k_i$ and $g_i$ being positive constants, can be found such that $\mathbf{A}_c=\mathbf{I}_{N}\otimes \mathbf{A}-\mathcal{L}_{\bar{\mathcal{G}}}\otimes(BK)$ and $\mathbf{A}_0=\mathbf{I}_N\otimes(\mathbf{A}-G^{\top}C) $ are Hurwitz, where $\otimes$ denotes the Kronecker product. 
%\end{lemma}

\subsection{Properties of two classes of matrix pencils}
We summarize the properties of the two forms of matrix pencils used in this paper into the following lemma. \begin{lemma}[{[\citen{2020-NYU-IFAC}]}]\label{lemma-matrix}
	Given real square matrices $\mathbf{Q}_1$ and $\mathbf{Q}_2$, the generalized eigenvalues of matrix pencil $\mathbf{Q}_1-s\mathbf{Q}_2$ are defined as the values of $s$ that make $\det(\mathbf{Q}_1-s\mathbf{Q}_2) = 0$. The set of generalized eigenvalues of matrix pencil $\mathbf{Q}_1-s\mathbf{Q}_2$ are denoted as $\sigma(\mathbf{Q}_1, \mathbf{Q}_2)$. 
	\begin{itemize}
		\item[1)] If $\mathbf{Q}_1$ is symmetric and $\mathbf{Q}_2$ is symmetric positive-definite (SPD), then $\mathbf{Q}_1-s\mathbf{Q}_2<0$ holds for all $s > \max\{\sigma(\mathbf{Q}_1, \mathbf{Q}_2)\}$;
		\item[2)] If  both $\mathbf{Q}_1$ and $\mathbf{Q}_2$ are symmetric negative-definite (SND), $\mathbf{Q}_1-s\mathbf{Q}_2< 0$ holds for all $s < \min\{\sigma(\mathbf{Q}_1, \mathbf{Q}_2)\}$.
	\end{itemize} 
\end{lemma} 

\textcolor{black}{In implementation,  the  set of the desired generalized eigenvalues  $\sigma\left(\mathbf{Q}_{1}, \mathbf{Q}_{2}\right)$ can be obtained through the code $s=\operatorname{double(solve(det(}\mathbf{Q}_{1}-s*\mathbf{Q}_{2})))$ with Matlab.}

\subsection{A Lyapunov-like function}
\textcolor{black}{In this paper, the time-derivative of $V(\tau)$ might be indefinite since the sign of $-\kappa_0+\kappa_1/\alpha(\tau)$ is not required to be negative for all $\tau$,  therefore the function $V (\tau)$ in Theorems 1-3 is not a Lyapunov function in the usual sense. For this special type of $V $, we need the following lemma to facilitate stability analysis.}
\begin{lemma}\label{lemma-2}
	Consider a function $V(\tau):[0,\infty)\rightarrow[0,\infty)$, if there exist two positive constants $\kappa_0$ and $\kappa_1$ such that
	\begin{equation}\label{lemma-V}
	\begin{aligned}
	\frac{dV}{d\tau}\leq- \left(\kappa_0 - \frac{\kappa_1}{\alpha(\tau)} \right) {V}, 
	\end{aligned}
	\end{equation}
	where $\alpha(\tau)=e^{\tau}/T$ is a monotonically increasing function to infinity as defined in (\ref{6}), $T$ is a positive design parameter. Then, it holds that  $V(\tau)\in\mathcal{L}_\infty[0,\infty)$ and  $\lim_{\tau\rightarrow \infty}V(\tau)=0$. 
\end{lemma}
\textit{Proof:}  
Since $\kappa_0$ is a positive constant and $\kappa_1/\alpha(\tau)$ is a monotonically decreasing function to zero, thus there must exist a time $\tau_1\in[0,\infty)$, such that, for all $\tau\in[\tau_1,\infty)$,
\begin{equation}\label{55}
\kappa_0-\frac{\kappa_1}{\alpha(\tau)}>0.
\end{equation}
Firstly, we analyze the boundedness of $V$ on $[0,\tau_1)$. From (\ref{lemma-V}) and the property of $\alpha(\tau)$, we know that
\begin{equation}\label{42-1}
\frac{dV}{d\tau}\leq \frac{\kappa_1}{\alpha(0)}V,
\end{equation}
where $\alpha(0)=1/T$, and then by solving (\ref{42-1}) it is easily to obtain 
\begin{equation}\label{57}
V(\tau)\leq e^{T\kappa_1 \tau_1}V(0)\in\mathcal{L}_\infty[0,\tau_1).
\end{equation}
Then, for $\tau\in[\tau_1,\infty)$, by means of (\ref{55}), Eq. (\ref{lemma-V}) satisfies
$
\frac{dV}{d\tau}\leq - {\kappa}V,$
where $\kappa=\kappa_0-\kappa_1/\alpha(\tau_1)$ is a positive constant, which yields
\begin{equation}\label{59}
V(\tau)\leq e^{-\kappa \tau}V(\tau_1) \in\mathcal{L}_\infty[\tau_1,\infty).
\end{equation}
Combining with (\ref{57}) and (\ref{59}), we obtain that 
\begin{equation*}
V(\tau)\leq e^{-\kappa \tau}e^{T\kappa_1 \tau_1}V(0),
\end{equation*}
which implies that $V(\tau)\in\mathcal{L}_\infty[0,\infty)$ and  $\lim_{\tau\rightarrow \infty}V(\tau)=0$. The proof is completed. $\hfill\blacksquare$
 
\balance

\end{document}